\documentclass[fleqn,usenatbib]{mnras}

\usepackage{newtxtext,newtxmath}
\usepackage{textcomp}
\usepackage[T1]{fontenc}
\usepackage{orcidlink}
\DeclareRobustCommand{\VAN}[3]{#2}
\let\VANthebibliography\thebibliography
\def\thebibliography{\DeclareRobustCommand{\VAN}[3]{##3}\VANthebibliography}

\usepackage{graphicx}	
\usepackage{amsmath}	
\usepackage{upgreek}
\usepackage{booktabs}
\usepackage{gensymb}
\usepackage{textcomp}
\graphicspath{{./}{image/}}
\usepackage{longtable}

\title[Metallicity gradients at $z\sim 0.3$]{The MAGPI Survey: forward modelled gas-phase metallicity gradients in galaxies at $z\sim 0.3$}

\author[Y. Mai et al.]{
Yifan Mai$^{\orcidlink{0000-0003-3514-6280}}$,$^{1,2,3,4}$\thanks{E-mail: yifan.mai@sydney.edu.au}
Scott M. Croom$^{\orcidlink{0000-0003-2880-9197}}$,$^{1,3}$
Emily Wisnioski$^{\orcidlink{0000-0003-1657-7878}}$,$^{3,5}$
Andrew J. Battisti$^{\orcidlink{0000-0003-4569-2285}}$,$^{6,3,5}$
J. Trevor Mendel$^{\orcidlink{0000-0002-6327-9147}}$,$^{3,5}$
\newauthor
Marcie Mun$^{\orcidlink{0000-0002-3706-9955}}$,$^{3,5}$
Caroline Foster$^{\orcidlink{0000-0003-0247-1204}}$,$^{3,7}$
Katherine E. Harborne$^{\orcidlink{0000-0002-2043-7985}}$,$^{8,9,10}$
Claudia D. P. Lagos$^{\orcidlink{0000-0003-3021-8564}}$,$^{3,6}$
Iris Breda$^{\orcidlink{0000-0002-3764-5780}}$,$^{11}$
\newauthor
Tianmu Gao$^{\orcidlink{0000-0002-1158-6372}}$,$^{3,5}$
Kathryn Grasha\thanks{ARC DECRA Fellow}$^{\orcidlink{0000-0002-3247-5321}}$,$^{3,5}$
Tamal Mukherjee$^{\orcidlink{0009-0004-7639-869X}}$,$^{4,12}$
Adriano Poci$^{\orcidlink{0000-0002-5422-7441}}$,$^{13}$
Rhea-Silvia Remus$^{\orcidlink{0009-0008-9260-7278}}$,$^{14}$
\newauthor
Piyush Sharda$^{\orcidlink{0000-0003-3347-7094}}$,$^{15}$
Sarah M. Sweet$^{\orcidlink{0000-0002-1576-2505}}$,$^{3,16}$
Sabine Thater$^{\orcidlink{0000-0003-1820-2041}}$,$^{11}$
Lucas M. Valenzuela$^{\orcidlink{0000-0002-7972-9675}}$,$^{14}$
Glenn van de Ven$^{\orcidlink{0000-0003-4546-7731}}$,$^{11}$
\newauthor
Tayyaba Zafar$^{\orcidlink{0000-0003-3935-7018}}$$^{12}$
and Bodo Ziegler$^{\orcidlink{0000-0003-2856-1080}}$$^{11}$
\\
$^{1}$Sydney Institute for Astronomy (SIfA), School of Physics, The University of Sydney, NSW 2006, Australia\\
$^{2}$Australian Astronomical Optics, Macquarie University, Sydney, NSW 2109, Australia\\
$^{3}$ARC Centre of Excellence for All Sky Astrophysics in 3 Dimensions (ASTRO 3D)\\
$^{4}$Astrophysics and Space Technologies Research Centre, Macquarie University, Sydney, NSW 2109, Australia\\
$^{5}$Research School of Astronomy and Astrophysics, Australian National University, Canberra, ACT 2611, Australia\\
$^{6}$International Centre for Radio Astronomy (ICRAR), M468, The University of Western Australia, 35 Stirling Highway, Crawley, WA 6009, Australia\\
$^{7}$School of Physics, University of New South Wales, Sydney, NSW 2052, Australia\\
$^{8}$Institute for Computational Cosmology, Durham University, South Road, Durham DH1 3LE, UK \\
$^{9}$Centre for Extragalactic Astronomy, Durham University, South Road, Durham DH1 3LE, UK \\
$^{10}$Department of Physics, Durham University, South Road, Durham DH1 3LE, UK \\
$^{11}$Department of Astrophysics, University of Vienna, T\"urkenschanzstra{\ss}e 17, 1180 Vienna, Austria\\
$^{12}$School of Mathematical and Physical Sciences, Macquarie University, NSW 2109, Australia\\
$^{13}$Sub-Department of Astrophysics, University of Oxford, Oxford, UK\\
$^{14}$Universitäts-Sternwarte, Fakultät für Physik, Ludwig-Maximilians-Universität München, Scheinerstr. 1, 81679 München, Germany\\
$^{15}$Leiden Observatory, Leiden University, PO Box 9513, NL-2300 RA Leiden, the Netherlands\\
$^{16}$School of Mathematics and Physics, University of Queensland, St Lucia, Queensland 4072, Australia\\
}

\date{Accepted XXX. Received YYY; in original form ZZZ}

\pubyear{\the\year{}}

\def \revise {}
\def \reviewer {}
\def \paperrevise {}
\def \teamreview {}
\def \mnras {}
\def \mnrass {}
\begin{document}
\label{firstpage}
\pagerange{\pageref{firstpage}--\pageref{lastpage}}
\maketitle

\begin{abstract}
We measure the seeing-deconvolved gas-phase metallicity gradients of 70 star-forming galaxies at $z\sim 0.3$ from the MAGPI survey and investigate their relationship with galaxy properties to understand the mechanisms that influence the distribution of metals and shape the evolution of the galaxies. We use a Bayesian modelling technique, \textsc{blobby3d}, which accounts for seeing effects (beam smearing) and can model the substructures of the flux distribution. The median metallicity gradient of our sample is $\nabla \mathrm{[O/H]}=-0.013^{+0.059}_{-0.033}$ dex/kpc. Among the galaxies in our sample, 32.9\% have negative metallicity gradients (2$\sigma$ significance), 10.0\% have positive gradients and 57.1\% have flat gradients. The $\nabla \mathrm{[O/H]}$-$M_*$ relation of the MAGPI galaxies generally agrees with theoretical predictions, where a combination of stellar feedback, gas transport, and accretion shapes the metallicity profile, with the dominant processes varying with galaxy mass. We find a positive correlation between $\nabla \mathrm{[O/H]}$ and gas velocity dispersion ($r=0.36$), indicating that stronger gas turbulence is associated with flatter or inverted metallicity gradients, likely due to enhanced gas mixing. Additionally, smaller galaxies tend to have flatter or positive gradients, suggesting that metal dilution by gas accretion or removal via feedback-driven winds may outweigh metal enrichment in small galaxies.

\end{abstract}




\section{Introduction}

\teamreview{Heavy elements, as the products of stellar nucleosynthesis, provide a powerful probe of star formation histories and galaxy evolution. The metal enrichment cycle in galaxies involves several interconnected processes. First, inflows of metal-poor gas from the intergalactic medium (IGM) provide the fresh fuel for star formation but can also dilute the existing metal content of the interstellar medium \citep[ISM; e.g.][]{Keres:2005,Grand:2019,Barbani:2025}. Star formation then generates new heavy elements, which are returned to the ISM through stellar winds and supernovae. Feedback from massive stars, supernovae, and active galactic nuclei (AGN) can subsequently drive outflows, expelling enriched material and redistributing metals within and beyond galaxies \citep{Muratov:2015,Mitchell:2020,Veilleux:2020,Wright:2024}. These competing processes together regulate the chemical evolution of galaxies \citep[e.g.][]{Peroux:2020,Sharda:2024}.}

The distribution of metals across galaxies therefore provides insight into different galaxy processes. The metallicity gradient of low-redshift galaxies is \revise{typically} negative \citep[i.e., more metals in the centre than in the outskirts;][]{Sanchez:2014,Belfiore:2017}, in agreement with an inside-out formation scenario \teamreview{where galactic discs have primarily formed by in-situ star formation}. \paperrevise{In the inside-out formation scenario, the core of a galaxy forms early while the disc builds up through gradual processes, such as accretion or minor mergers \citep{Larson:1976, Kepner:1999,Portinari:1999,Sanchez:2014,Baker:2025}.} 
\teamreview{Metallicity gradients are also shaped by gas accretion, gas transportation, galactic wind and galaxy interactions \citep{Cresci:2010,Kewley:2010,Rupke:2010,Marinacci:2014,Chisholm:2018}. These mechanisms can induce gas mixing, metallicity dilution and redistribution, leading to the flattening or even inversion of gradients. Therefore, measuring the metallicity gradient can offer insights into the interplay between galaxy growth, feedback, and environmental effects \citep{Mayor:1981,Pezzulli:2016,Collacchioni2020}.} 

It is not clear which mechanisms dominate in setting metallicity gradients as a function of cosmic time. 
%
Simulations do not agree on the relative importance of key parameters that shape metallicity gradients in part \mnras{due to different treatments of star formation \citep{Chiappini:2001,Fu:2009,Pilkington:2012}, feedback \citep{Kobayashi:2011,Gibson:2013,Ma:2017,Yates:2021,Helmer:2021}, and accretion \citep{Molla:2005, Ma:2017,Tissera:2022,Sun:2024}.}
%
Observations of \revise{high-redshift} galaxies provide constraints on metallicity gradient evolution \citep{Queyrel:2012,WangXin:2022,Venturi:2024,Ju:2025,Li:2025highredshift}. However, compilations of observational results across $z \sim 0$–4 \citep[e.g.][]{Curti:2020} reveal substantial scatter, with most measurements consistent with no significant gradient, while some exhibit positive gradients.
%
%
If confirmed, the variation of metallicity gradients at high redshift ($z \gtrsim 1$) suggests the \paperrevise{relative} importance of mechanisms \paperrevise{may be} different to local galaxies. Positive gradients, which are rare at low redshift, may result from strong gas accretion that dilutes the metallicity at the galaxy centre \citep[e.g.][]{Cresci:2010,Queyrel:2012,Stott:2014}. \mnras{The flatter gradients found at high redshift may also result from more regular metal-poor gas inflows \citep{Sun:2024}, strong feedback \citep{Ma:2017,Sun:2024}, or mergers \citep{Troncoso:2014,Venturi:2024,Vallini:2024}}. \mnras{Studies of gas kinematics have shown that, on average, the gas turbulence in higher-redshift galaxies is higher than at lower redshift} \citep{Kassin:2012, Wisnioski:2015,Ubler:2019,Mai:2024}. \mnras{However, recent observations have detected dynamically cold, rotationally supported discs at high redshift \citep{Rizzo:2021,Parlanti:2023}, suggesting that galaxy mass may play an important role in maintaining such discs \citep{Gurvich:2023, Kohandel:2024}.} This suggests that the large scatter of metallicity gradients at high redshift could be connected, in part, to strong gas mixing driven by stellar feedback as well as gas transport and accretion, \mnras{and that the efficiency of these processes may vary with galaxy mass} \citep{Sharda:2021b, Sharda:2024}.

However, making consistent measurements of metallicity gradients across cosmic time is challenging. Some amount of scatter and flattening amongst gradient measurements are likely due to observational effects and \teamreview{measurement} biases. For example, \paperrevise{different observational studies report varying degrees of scatter in the metallicity gradient in high redshift galaxies \citep[e.g.][]{Wuyts:2016,Carton:2018,Curti:2020,Gillman:2021}. Systematic differences in sample selection (e.g. stellar mass range) and measurement methods, such as the choice of metallicity diagnostic \citep{Poetrodjojo:2021} and the treatment of seeing, or beam smearing, likely also contribute to the observed scatter.}
\paperrevise{
\teamreview{Beam smearing flattens both positive and negative intrinsic metallicity gradients, with the degree of flattening increasing at lower angular resolution, i.e., larger point-spread function (PSF), and for smaller or more inclined galaxies} \citep{Yuan:2013,Stott:2014,Belfiore:2017,Wuyts:2016,Acharyya:2020}. These constraints are particularly relevant at higher redshifts where observations are often limited by angular resolution and surface brightness.

} 
%
%
%
The measurement of metallicity gradients is also influenced by the clumpiness of gas \teamreview{\citep{Belfiore:2017,Poetrodjojo:2019,Metha:2021,Metha:2022}}. \teamreview{Clumpy gas can lead to enhanced star formation in local areas, which increases the intrinsic scatter of the metallicity distribution.} \paperrevise{The inferred metallicity of clumpy galaxies, especially at low spatial resolution, may be \revise{biased} towards bright H\,\textsc{ii} regions, affecting the overall gradient \citep[e.g.][]{Carton:2017}.} \revise{Models} that assume \revise{a smooth radial flux distribution} cannot fully capture the spatial variance of \revise{the} metallicity distribution for galaxies with complex substructures. 

In fact, most published metallicity gradients do not take beam smearing into account when modelling the flux distribution. Some studies have applied generalised corrections based on simple smooth models \citep[e.g.][]{Wuyts:2016}. Uniquely, \cite{Carton:2017} presented a forward modelling approach. Here we build on the results of \cite{Carton:2017} to obtain a seeing-deconvolved metallicity distribution \textit{which captures the substructures in emission line flux}. This requires a modelling technique with more flexibility in reproducing realistic flux distributions. We use \textsc{blobby3d} \citep{Varidel:2019}, which employs a forward modelling technique that accounts for beam smearing and models the gas kinematics and spatial substructures simultaneously. \mnras{This approach allows us to analyse the relationship between metallicity distribution and seeing-deconvolved gas turbulence, which can also be affected by beam smearing \citep{Davies:2011,DiTeodoro:2015,Bouche:2015,Kohandel:2020, Rizzo:2022, Mai:2024}.}




In this study, we measure the gas-phase metallicity gradients of star-forming galaxies and aim \revise{to} understand the mechanisms related to the metallicity distribution at $z\sim 0.3$. \teamreview{This redshift represents a key transitional epoch between the well-studied local and high-redshift Universe, providing critical insights into the evolution of gas kinematics \citep[e.g.][]{Wisnioski:2015,Mai:2024} and metallicities \citep[e.g.][]{Sharda:2021d}. \mnras{Moreover, the high spatial resolution achieved in the MAGPI observations (PSF $\approx 0.6''$, corresponding to $\approx2.7$\ kpc at $z=0.3$) allows us to probe metallicity gradients at spatial scales comparable to those of local Integral Field Unit (IFU) surveys (e.g. the MaNGA survey; \citealt{Bundy:2015}, with a PSF of $\approx 2.5''$ corresponding to $\approx 2.4$\ kpc at $z=0.05$), thereby enabling direct comparisons across cosmic time.}} 

We arrange the rest of the paper as follows: In Section \ref{sec:data-4}, we introduce the MAGPI survey and our sample selection. In Section \ref{sec:methods-4}, we introduce the method we use to model the flux distribution and calculate the metallicity and metallicity gradients. In Section \ref{sec:results-4}, we demonstrate the results for the metallicity gradients of the star-forming MAGPI galaxies and their correlation with other galaxy properties. In Section \ref{sec:discussion-4}, we discuss the evolution of metallicity gradients and the mechanisms that relate to metallicity distribution. In Section \ref{sec:conclusions-4}, we summarise our results.

Throughout this paper, we assume a \citet{Chabrier:2003} stellar initial mass function (IMF) and adopt \revise{a} concordance cosmology ($\Omega_{\Lambda} = 0.7$, $\Omega_{m} = 0.3$, $H_0 = 70\: \mathrm{km\: s^{-1}\: Mpc^{-1}}$).


\section{Data}\label{sec:data-4}

\subsection{The MAGPI survey}

The Middle Ages Galaxy Properties with Integral
field spectroscopy (MAGPI) survey \citep{Foster:2021} focuses on galaxies at a lookback time of $3-4$ Gyr ($z\sim 0.3$). The survey uses the Multi Unit Spectroscopic Explorer \citep[MUSE;][]{Bacon:2010} with Ground Layer Adaptive Optics (GLAO), mounted on the UT4 of the Very Large Telescope (VLT). MAGPI uses the wide-field mode of MUSE, which has a 1 $\times$ 1 arcmin$^{2}$ field-of-view (FOV) sampled at 0.2 $\times$ 0.2 arcsec$^{2}$. The typical full width at half maximum (FWHM) of the \teamreview{MAGPI observations} PSF is $\sim$0.6\,arcsec. The MAGPI survey covers a wavelength range between 4700 and 9351 \AA\ with a spectral resolution of $R = 2728$ at 7025 \AA.

The MAGPI survey targets 60 massive central galaxies (primary galaxies) and their neighbours (secondary galaxies) within the MUSE FOV. The selection of MAGPI primary galaxies \revise{required} that galaxies uniformly sample a range of environments (isolated, groups and clusters), covering a halo mass range $11.35 \leq \textup{log} (\textup{M}_\textup{halo}/\textup{M}_{\odot})\leq 15.35$. \mnras{The halo masses are taken from the GAMA group catalogue \citep{Robotham:2011}, where group halo masses are estimated by dynamical masses based on group size and velocity dispersion, and then calibrated against mock galaxy catalogues derived from $N$-body simulations \citep[see Section 4.3 in][]{Robotham:2011}.} 56 primary galaxies \revise{were} selected from the Galaxy And Mass Assembly survey \citep[GAMA;][]{Driver:2011} G12, G15 and G23 fields. The remaining 4 primary galaxies \teamreview{were obtained} from MUSE archival observations of Abell 370 and Abell 2744. At the time of writing, the survey has completed the observation of all 56 fields, which have been fully reduced with relevant data products. \teamreview{However, only 53 fields selected from the GAMA survey are used in this study, as the observation and reduction of the remaining fields were completed after the analysis presented here was finalised.} There are \reviewer{332} star-forming galaxies that have H$\alpha$-based star formation rate (SFR) measurements at redshift $0.26<z<0.42$.


We utilise the Galaxy IFU Spectroscopy Tool \citep[\textsc{GIST};][]{Bittner:2019} to fit \revise{the observed (seeing-convolved)} emission lines and continuum spectra in MAGPI galaxies (Battisti et al. in preparation). \mnras{\textsc{GIST} is a Python wrapper that implements the Penalized PiXel-Fitting \citep[\textsc{pPXF};][]{Cappellari:2004,Cappellari:2017} and the "Gas and Absorption Line Fitting" \citep[\textsc{GandALF};][]{Sarzi:2006,Falcon:2006} method on integral field spectroscopy (IFS) data. The stellar continuum fits are conducted on a spaxel level. The stellar continuum is fit using the \textsc{SSP\_MIST\_C3K\_Salpeter} templates \citep{Conroy:2009,Conroy:2010a,Conroy:2010b}, which are based on the Modules for Experiments in Stellar Astrophysics (\textsc{MESA}) Isochrones and Stellar Tracks \citep[\textsc{MIST};][]{Choi:2016}, and the lines are masked at that step. The stellar continuum models span a grid of ages and metallicities, comprising 107 age steps covering $\log(\textup{age/yr}) = 5.0$–$10.3$ in increments of 0.05, and 12 metallicity steps spanning $\log(Z/Z_\odot) = [-2.5,0.5]$, where $Z_\odot$ is solar metallicity. The emission lines are fit using single-component Gaussian functions where all line kinematics are tied together.}





The stellar masses are determined using \textsc{ProSpect} \citep{Robotham:2020}, which is an energy balance spectral energy distribution (SED) fitting code. Photometry from GAMA imaging over the $ugriZYJHK$ bands is employed for the \textsc{ProSpect} SED fitting, assuming a \citet{Chabrier:2003} IMF. \mnrass{We note that the photometric data have a wider spectral coverage into the near-infrared than the MAGPI data, providing more robust estimates for the stellar mass.} Stellar population templates from \citet{Bruzual:2003} are utilised to create an unattenuated stellar SED, and the dust attenuation law by \citet{Charlot:2000} is applied to correct for dust attenuation, \mnras{following the methodology described in Section 3 of \citet{Bellstedt:2020b}.} \teamreview{We use the tool \textsc{GALFIT} \citep{Peng:2002} to estimate the effective radius ($R_\textup{e}$) of the galaxies from synthetic $i$-band images generated from the MUSE data.} \mnras{\textsc{GALFIT} accounts for the PSF and models the 2-dimensional surface brightness distribution assuming a single Sérsic component \citep{Sersic:1963}.} The stellar mass surface density ($\Sigma_{M_*}$) is calculated as $\Sigma_{M_*} = M_*/(2\pi R_\mathrm{e}^2)$.

The ionised gas velocity dispersion ($\sigma_\textup{gas}$), which traces the strength of gas turbulence, is measured using \textsc{blobby3d} and taken from \citet{Mai:2024}. \mnras{This value includes both the turbulent and thermal components of the gas motion.} Beam smearing due to the atmosphere \reviewer{broadens} the observed gas velocity dispersion, especially for \reviewer{galaxies} with a \reviewer{steep} velocity gradient. \textsc{blobby3d} can measure the intrinsic $\sigma_\textup{gas}$ by forward-modelling the gas kinematics and spatial gas distribution simultaneously accounting for beam smearing from the PSF. Additionally, we extract the rotational velocity at one effective radius ($v_\phi$) for each galaxy from the same \textsc{blobby3d} models. \mnras{The details of \textsc{blobby3d} are introduced in Section \ref{sec:methods-4}.} 

We use the dust-corrected H$\alpha$ flux maps to derive SFR using the calibration of \citet{Kennicutt:1998}. We correct the dust extinction using the Balmer decrement on a spaxel by spaxel basis \citep[see details \mnras{in Section} \ref{sec:extinctioncorrection} and][]{Mun:2024}. The extinction curve from \citet{Fitzpatrick:2019} is adopted with $R_V = 3.1$. The SFR surface density ($\Sigma_{\rm SFR}$) is calculated as $\Sigma_{\rm SFR} = \mathrm{SFR}/(2\pi R_{\mathrm{e,H}\alpha}^2)$, in which \revise{SFR is the total integrated SFR}, and $R_{\mathrm{e,H}\alpha}$ is the half-light radius of the H$\alpha$ map, which is used to better trace the star formation region. The star formation main sequence (SFMS) is defined as the linear best fit to the MAGPI galaxies at $0.25 \leq z \leq 0.424$ \citep{Mun:2024}.

\subsection{Sample selection}

\begin{figure}
\centering
    \includegraphics [width=0.45\textwidth]{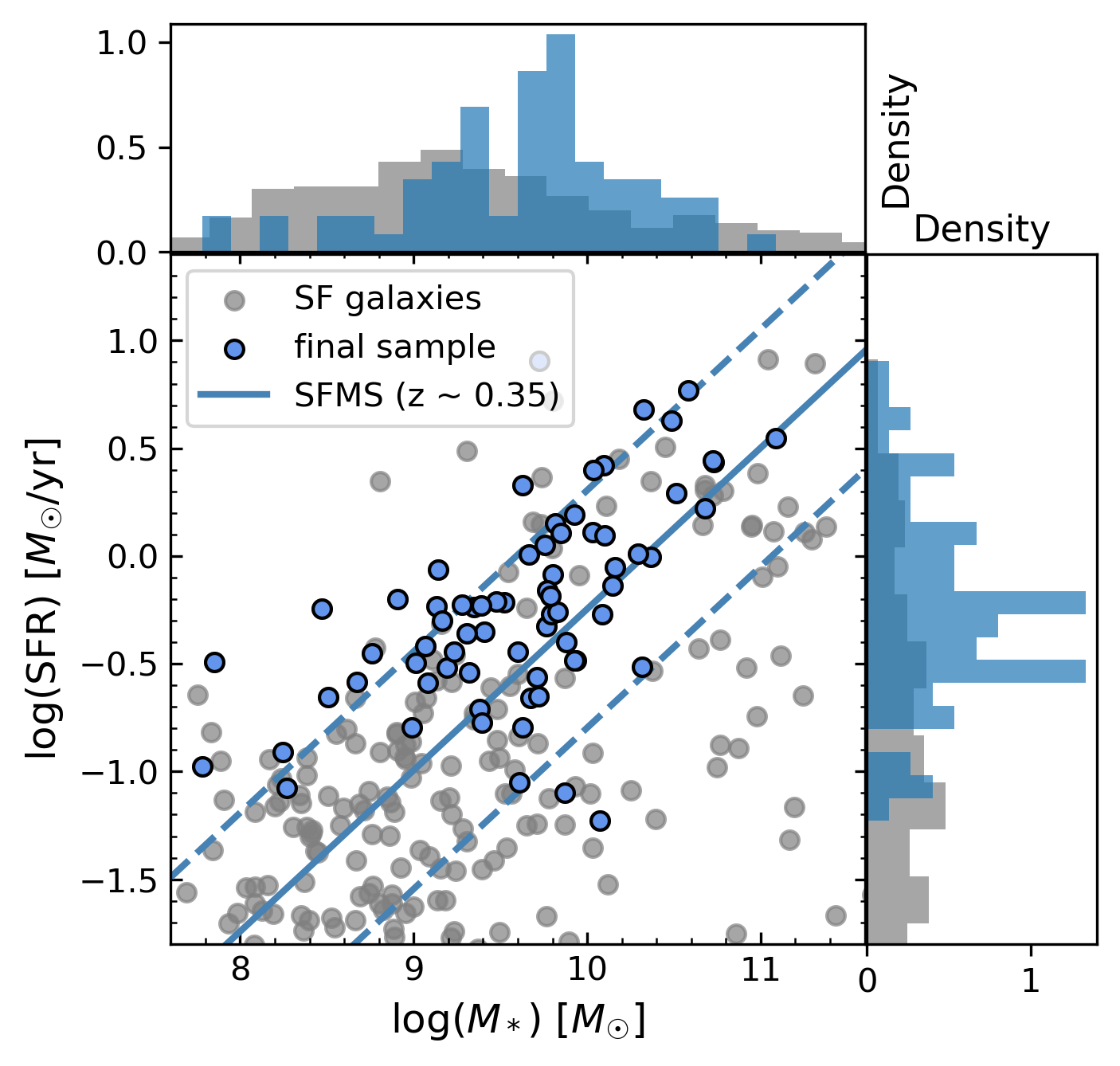}
    \caption{\teamreview{The distribution of H$\alpha$-based SFR and $M_*$ for all star-forming galaxies and galaxies in our final sample. The solid line represents the SFMS for galaxies at $z\sim 0.35$, with the dashed lines representing the root-mean-square error of SFMS \citep{Mun:2024}.}}
    \label{fig:mass-SFR}
\end{figure}

We select galaxies in the redshift range $0.26<z<0.42$, ensuring that the emission lines we use for gas-phase metallicity diagnostics and dust extinction correction fall within the wavelength coverage of the MAGPI survey. These emission lines include [O\,\textsc{ii}]$\uplambda 3727$, $\uplambda 3729$, H$\beta$, H$\alpha$ and [N\,\textsc{ii}]$\uplambda 6584$. We remove galaxies with evidence of active galactic nucleus (AGN) activity to avoid AGN contamination of the metallicity measurements. We use the Baldwin, Phillips \& Terlevich \citep[BPT;][]{baldwin:1981,Kewley:2001,Kauffmann:2003} ionisation diagnostic diagrams to classify spaxels as star-forming (SF), AGN and composite. \teamreview{Only spaxels with a S/N greater than 3 in all relevant emission lines are included in the BPT classification.} \mnras{Galaxies hosting AGN or composite central regions were excluded based on spatially resolved BPT classification maps. Specifically, we removed galaxies where more than five contiguous central spaxels (corresponding to one PSF area) fall within the AGN or composite region of the BPT diagram. Galaxies with only isolated AGN/composite spaxels were retained, as such cases are likely due to noise. We note that this cut removes the most massive galaxies ($M_* \sim 10^{11}\ \textup{M}_\odot$), which more likely have central AGN/composite region.} We select galaxies with metallicity measurements extending to at least twice the PSF FWHM (PSF$_\mathrm{FWHM}$), ensuring robust metallicity gradient estimates. \mnras{We note that this criterion preferentially removes galaxies with low SFR that lack sufficiently extended line emission at large radii.} There are 70 galaxies in our sample after applying the above cuts \teamreview{(full list in Appendix \ref{sec:compiled data})}. \teamreview{We show the distribution of SFR and $M_*$ for galaxies in our final sample in Figure \ref{fig:mass-SFR}.}

\section{Methods}\label{sec:methods-4}

We use \textsc{blobby3d}\footnote{https://github.com/SpaceOdyssey/Blobby3D} \citep{Varidel:2019} to model the spatial flux distribution of the emission lines we used for metallicity diagnostics. \textsc{blobby3d} is designed to perform Bayesian inference on gas kinematics from emission line observations of galaxies using integral field spectroscopy. It can model the gas spatial substructure and kinematics simultaneously, accounting for the PSF and line spread function (LSF), and assuming a continuous velocity profile defined by the \cite{Courteau:1997} empirical model \mnras{and a constant velocity dispersion across the disc}. \teamreview{\textsc{blobby3d} models the gas distribution as a sum of Gaussian components (blobs), without assuming a pre-defined flux distribution profile for the galaxy. The number of blobs is flexible \mnras{(typically in the range of $\sim1$–$300$)}, and the position, shape and flux of each blob are also flexible and depend on the complexity of the observed gas substructure.} This method can model complex gas substructures such as rings, spiral arms and individual clumps, which are difficult to model using methods that assume a parametric radial flux distribution. \mnras{For example, $^{\textup{3D}}$\textsc{BAROLO} \citep{DiTeodoro:2015} adopts a concentric tilted-ring model, in which the galaxy is represented by a set of co-centric rings with varying flux. This approach is less flexible than \textsc{Blobby3D}. However, $^{\textup{3D}}$\textsc{BAROLO} may perform more robustly on data with lower S/N due to its simpler parametric framework.}

\textsc{blobby3d} uses a diffusive nested sampling algorithm, \textsc{DNest4} \citep{Brewer:2011,Brewer:2018}, which is an effective variant of nested sampling  \citep{Skilling:2004}. \textsc{DNest4} uses Markov Chain Monte Carlo (MCMC) to explore the complex posterior distribution and compute the normalization constant. The key advantage of \textsc{DNest4} over other sampling algorithms is its effectiveness in handling high-dimensional parameter spaces and multimodal distributions, which is a useful feature for modelling complex gas substructures.

An inference problem with a varying number of components is normally difficult to explore. \textsc{DNest4} has an in-built reversible jump object that allows variations to the number of blobs/components. \textsc{blobby3d} uses the reversible jump to add or remove blobs from the model, which enables the number of blobs to be a flexible parameter depending on the complexity of the galaxy flux distribution. See Table 1 in \citet{Varidel:2019} for more details about all hyperparameters and parameters used in the model.

\subsection{Metallicity diagnostics}

We use \revise{the} [N\,\textsc{ii}]/[O\,\textsc{ii}] ratio (N2O2), based on [N\,\textsc{ii}]$\uplambda 6584$ and [O\,\textsc{ii}]$\uplambda \uplambda 3727,\ 3729$ lines, as our primary gas-phase metallicity diagnostic. N2O2 is the most reliable metallicity diagnostic in the optical spectrum as it is less sensitive to ionisation parameter and the ISM pressure. Note that N2O2 can be more sensitive to the ionisation parameter at lower metallicities \citep[see \revise{the} review by][]{Kewley:2019}.

We use the metallicity diagnostic calibration outlined in \citet{Kewley:2019}. The gas-phase metallicity is calculated as

\begin{equation}
\begin{aligned}
    12+\log(\text{O/H}) = {}& 9.4772 + 1.1797 \cdot  x + 0.5085\cdot y + 0.6879\cdot x\cdot y \\
     & + 0.2807\cdot x^2 
            + 0.1612\cdot y^2 + 0.1187\cdot x\cdot y^2  \\
            & + 0.1200\cdot y\cdot x^2 + 0.2293\cdot x^3
                + 0.0164\cdot y^3~,
\end{aligned}
\end{equation}

\noindent where $12+\textup{log(O/H)}$ is the gas-phase metallicity, $x$ is the metallicity diagnostic ratio 
\[
x = \log\left( \frac{F_\text{int}([\text{N\,\textsc{ii}}]\,\uplambda6584)}{F_\text{int}([\text{O\,\textsc{ii}}]\,\uplambda3727) + F_\text{int}([\text{O\,\textsc{ii}}]\,\uplambda3729)} \right)
\]
for the N2O2 \reviewer{diagnostic}. \reviewer{$F_{\textup{int}}$ is the intrinsic flux after dust extinction correction of a given emission line (see Section \ref{sec:extinctioncorrection} for more details).} $y$ is the dimensionless ionisation parameter $\textup{log}(U)$. Here we adopt log$(U)=-3.17$. \paperrevise{We note that a reasonable range of values \teamreview{($-3.98<\textup{log}(U)<-1.98$) makes little difference \citep{Kewley:2019}.}}

We use \revise{the} [N\,\textsc{ii}]$\uplambda 6584$/H$\alpha$ ratio (N2H$\alpha$) to measure the metallicity as a comparison \citep[P04;][]{Pettini:2004}. The advantage of this diagnostic is that the separation between [N\,\textsc{ii}] and H$\alpha$ emission lines is small, thus there is no need to make a dust extinction correction for the line ratio calculation, which may introduce more uncertainty, especially when the H$\beta$ line is weak. \paperrevise{It is also an advantage for observations at $z>0.5$ when obtaining multiple optical line groups from ground-based facilities is often time prohibitive. However, the N2H$\alpha$ diagnostic is \textit{more} sensitive to the ionisation parameter and ISM pressure \teamreview{than the N2O2 diagnostic \citep{Kewley:2019}}.} The gas-phase metallicity using N2H$\alpha$ is calculated as 

\begin{equation}
    \begin{aligned}
        12+\log(\text{O/H}) = {} & 9.37 + 2.03 \cdot x + 1.26\cdot x^2 + 0.32\cdot x^3 ~,
    \end{aligned}
    \label{eq:n2ha_p04}
\end{equation}

\noindent where 
\[
x = \log\left( \frac{F_\textup{int}(\text{[N\,\textsc{ii}]}\,\uplambda6584)}
{F_\textup{int}(\text{H}\,\alpha)} \right).
\]

\noindent \mnrass{The P04 calibration is derived from direct $T_e$-based abundances and therefore does not explicitly depend on the ionisation parameter; instead, it assumes that the ionisation conditions of the target H\,\textsc{ii} regions are similar to those of the calibration sample.}

\teamreview{We also measure the N2H$\alpha$ metallicity using the calibration functional form from \citet{Kewley:2019} (K19). We adopt the functional form from P04 for our main analysis, as it has a larger valid range (valid over $7.17 < 12+\log(\text{O/H})<8.86$) compared to the K19 calibration (valid over $7.6 < 12+\log(\text{O/H})<8.53$). The average difference in metallicity gradients between N2H$\alpha_\mathrm{P04}$ and N2H$\alpha_\mathrm{K19}$ in our sample is only 0.0016 dex/kpc. Therefore, the choice of calibration functional form does not affect our conclusions.}

\subsection{Modelling the flux distribution}\label{sec:model_flux_distribution}


\reviewer{\textsc{blobby3d} requires a single, continuous wavelength range for input spectra. It does not support the inclusion of multiple disjoint wavelength intervals, meaning that any analysis requiring non-contiguous spectral regions must be performed separately. \textsc{blobby3d} is computationally expensive, making it impractical to fit a large wavelength range that simultaneously includes H$\alpha$, [N\,\textsc{ii}], and [O\,\textsc{ii}]. To optimise computational efficiency, we first fit the spectrum covering H$\alpha$ and [N\,\textsc{ii}], as these lines are close in wavelength. Although [N\,\textsc{ii}] is relatively weak, H$\alpha$ provides a strong constraint on the kinematics. We then fit a separate spectral range that includes [O\,\textsc{ii}].}

To model the flux of H$\alpha$ and [N\,\textsc{ii}], we select the rest-frame wavelength range [6535, 6595]\,\AA\  of the cube, which covers H$\alpha$ and [N\,\textsc{ii}]$\uplambda \uplambda 6548,\ 6584$. We subtract the stellar continuum from the spectrum using the output from \textsc{gist}, as \textsc{blobby3d} requires the input datacube to be continuum subtracted. \reviewer{The blobs for H$\alpha$ and [N\,\textsc{ii}] \mnras{have the same shape and position}.} \textsc{blobby3d} performs inferences for the H$\alpha$ flux and the [N\,\textsc{ii}]/H$\alpha$ flux ratio for each blob (spatial Gaussian flux profile). The flux ratio between the two [N\,\textsc{ii}] lines is assumed to be $F_{6584}/F_{6548}=3$ \citep{Acker:1989}. \paperrevise{We tie together the velocity and dispersion of H$\alpha$ and both [N\,\textsc{ii}] lines, such that all three lines share the same kinematic parameters.}

We then model the flux of [O\,\textsc{ii}]$\uplambda \uplambda 3727,\ 3729$ using the data in the rest-frame wavelength range [3718, 3738]\,\AA. The flux ratio between [O\,\textsc{ii}]$\uplambda 3727$ and [O\,\textsc{ii}]$\uplambda 3729$ is set to be flexible for each blob. We constrain the kinematic model of [O\,\textsc{ii}] to have the same position angle (PA) \revise{as} H$\alpha$ using the average PA we obtain from the inference of H$\alpha$. This PA constraint is under the assumption that all gas components rotate in the same direction. \paperrevise{The flux distribution (the number, position and shape of blobs) of [O\,\textsc{ii}] is modelled independently from H$\alpha$, however we notice that in practice the two lines show similar \mnras{spatial} flux distributions in the majority of galaxies.}

\paperrevise{We calculate the standard deviation of the flux from the posterior effective samples of \textsc{blobby3d} and use it as flux uncertainty. The uncertainty in metallicity is then derived from the flux uncertainty through error propagation.}

\reviewer{\textsc{blobby3d} requires the FWHM of PSF and LSF as inputs. \paperrevise{The PSF for the MAGPI survey is reconstructed from information generated by the adaptive optics system \citep[][Mendel et al. in preparation]{Fusco:2020}. The LSF is measured by fitting Gaussian profiles to isolated bright sky lines \citep[][Mendel et al. in preparation]{Bacon:2017}.} We use the PSF of the \teamreview{MUSE-derived} $Z$-band when we model the H$\alpha$ and [N\,\textsc{ii}] lines, as the $Z$-band is close to the wavelength of H$\alpha$ at $z\sim 0.3$. We use the PSF of the \teamreview{MUSE-derived} $g$-band when we model the [O\,\textsc{ii}] lines. We adopt the FWHM of the LSF at $6563\cdot (1+z)$\,\AA \, when we model the H$\alpha$ and [N\,\textsc{ii}] lines, where $z$ is the redshift of the galaxy. We adopt the FWHM of LSF at $3728\cdot(1+z)$\,\AA\, when we model the [O\,\textsc{ii}] lines.}

We also test other levels of \reviewer{constraints} on the kinematics for [O\,\textsc{ii}] modelling, including (1) no \reviewer{constraints} on the kinematics; \mnrass{(2) using the posterior of the PA from H$\alpha$ as a prior for the PA from [O\,\textsc{ii}];} (3) constraining both \revise{the} PA and the velocity profile of [O\,\textsc{ii}] to be the same as H$\alpha$; (4) constraining the PA, velocity profile and velocity dispersion to be the same as H$\alpha$. The results show that the [O\,\textsc{ii}] flux distribution does not vary significantly between different constraints and they do not change the results on the metallicity gradient. Previous studies suggested that the velocity and the velocity dispersion measured from different emission lines may be different \citep[e.g.][]{Levy:2018,Law:2022,Thater:2022,Ubler:2024}. We therefore adopt the simplest constraint, constraining the PA from [O\,\textsc{ii}] the same as that from H$\alpha$, to improve our modelling and allow the flexibility of the velocity profile and velocity dispersion.

\reviewer{Different kinematics for various emission lines may indicate that they are not emitted from the same regions, which may influence the accuracy of the metallicity measurement. Therefore, measuring the metallicity using multiple diagnostics is crucial for assessing systematic differences and ensuring consistency across different methods.}


\subsection{Extinction correction}\label{sec:extinctioncorrection}

The flux of emission lines is attenuated by the dust in the ISM in the Milky Way and the source galaxy. The strength of dust attenuation is wavelength dependent, which causes a biased flux ratio if two emission lines are widely separated.  To get the intrinsic flux of emission lines, we perform two steps of dust attenuation correction, to correct the extinction from the Milky Way and the source galaxy using \revise{the} \textsc{dust\_extinction} Python package \citep{Gordon:2024}. 

Firstly, we correct the foreground extinction from the dust in the Milky Way using

\begin{equation}
    F_{\textup{MWcorr}}\left ( \uplambda_{\textup{red}}  \right )=F_{\textup{obs}}\left ( \uplambda_{\textup{red}}  \right )\cdot 10^{0.4\cdot k(\uplambda_{\textup{red}} )\cdot E(B-V)_{\textup{MW}}}~,
\end{equation}

\noindent where $F_{\textup{MWcorr}}$ is the flux of a given emission line after \revise{the} Milky Way dust extinction correction, $\uplambda_{\textup{red}}$ is the redshifted wavelength of the emission line, $F_{\textup{obs}}$ is the observed flux of the emission line, $k$ is the extinction curve from \citet{Fitzpatrick:2019} with $R_V=3.1$, and $E(B-V)_{\textup{MW}}$ is the reddening value due to the dust in Milky Way \citep{Schlafly:2011}. 

Then we correct the intrinsic extinction due to the dust from the source galaxy using 

\begin{equation}
    F_{\textup{int}}\left ( \uplambda_{\textup{res}}  \right )=F_{\textup{MWcorr}}\left ( \uplambda_{\textup{res}}  \right )\cdot 10^{0.4\cdot k(\uplambda_{\textup{res}} )\cdot E(B-V)_{\textup{int}}}~,
\end{equation}

\noindent where $F_{\textup{int}}$ is the intrinsic flux of a given emission line, $\uplambda_{\textup{res}}$ is the rest-frame wavelength of the emission line, and $E(B-V)_{\textup{int}}$ is the intrinsic reddening value due to the dust in the source galaxy. We calculate $E(B-V)_{\textup{int}}$ for each spaxel, as follows,

\begin{equation} \label{eq:e(b-v)}
    E\left ( B\!-\!V \right )_{\textup{int}} =
\frac{
\log_{10}\left ( \frac{\left ( \textup{H}\alpha/\textup{H}\beta \right )_{\textup{MWcorr}}}{\left ( \textup{H}\alpha/\textup{H}\beta \right )_{\textup{int}}} \right )
}{
\left( 0.4 \left( k(\textup{H}\beta) - k(\textup{H}\alpha) \right) \right)
}~,
\end{equation}

\noindent where $\left ( \textup{H}\alpha/\textup{H}\beta \right )_{\textup{MWcorr}}$ is the ratio of observed H$\alpha$ and H$\beta$ flux (Balmer decrement) after foreground extinction correction, $\left ( \textup{H}\alpha/\textup{H}\beta \right )_{\textup{int}}$ is the intrinsic ratio of H$\alpha$ and H$\beta$ and we adopt $\left ( \textup{H}\alpha/\textup{H}\beta \right )_{\textup{int}}=2.86$ for case B recombination \citep{Osterbrock:1989}. $k(\textup{H}\beta)$ and $k(\textup{H}\alpha)$ are the $k$ value from the extinction curve from \citet{Fitzpatrick:2019} for H$\alpha$ and H$\beta$, respectively.

We calculate the Balmer decrement using the H$\alpha$ and H$\beta$ flux measurement from the \textsc{gist} output. \mnrass{We note that it would be better to implement the Balmer decrement into the prior in \textsc{blobby3d} to take the \revise{seeing-deconvolved} structure of dust into account. However, such an implementation involves complex modifications of \textsc{blobby3d} and a detailed investigation of dust attenuation effects, which is beyond the scope of this paper}.

\subsection{Metallicity gradient measurement}

\mnras{For the N2O2 diagnostic, [N\textsc{ii}] and [O\textsc{ii}] are modelled independently due to their large wavelength separation. Because the recovered blobs are not in one-to-one correspondence, we construct a seeing-deconvolved flux map by summing the flux from all blobs for each emission line.} \teamreview{We measure the metallicity of each pixel and calculate the average metallicity as a function of radius, by binning the pixels in elliptical annuli with a bin width of 0.5 PSF$_\mathrm{FWHM}$.} 
We include only the pixels with S/N $> 3$ in the [N\textsc{ii}], [O\textsc{ii}], H$\alpha$, and H$\beta$ emission lines to ensure a reliable metallicity measurement in each pixel. 
When we calculate the average metallicity in each bin, three different weight methods are used: (1) unweighted: the metallicity of each pixel is weighted equally; (2) H$\alpha$ weighted: the metallicity is weighted by the H$\alpha$ flux at that pixel; (3) inverse variance weighted: the metallicity is weighted by the inverse variance of the metallicity at that pixels. The H$\alpha$ weighted method can lower the weights on the pixels that have low flux, however it can potentially bias the average metallicity. Similarly, the inverse variance weighted method can increase the weights of the pixels with more accurate metallicity measurements. 

\teamreview{We then fit the resulting metallicity–radius relation using ordinary least squares linear regression, implemented via the \texttt{curve\_fit} function from the \texttt{SciPy} library \citep{Virtanen:2020}, to obtain both the metallicity gradient and central metallicity. We normalised the gradient by the $R_\textup{e}$ of the galaxy to remove the size dependence of metallicity gradients \citep{Sanchez:2014,Ho:2015,Sanchez-menguiano:2016,Poetrodjojo:2021}.}

\begin{figure*}
    \centering
    \includegraphics [width=0.85\textwidth]{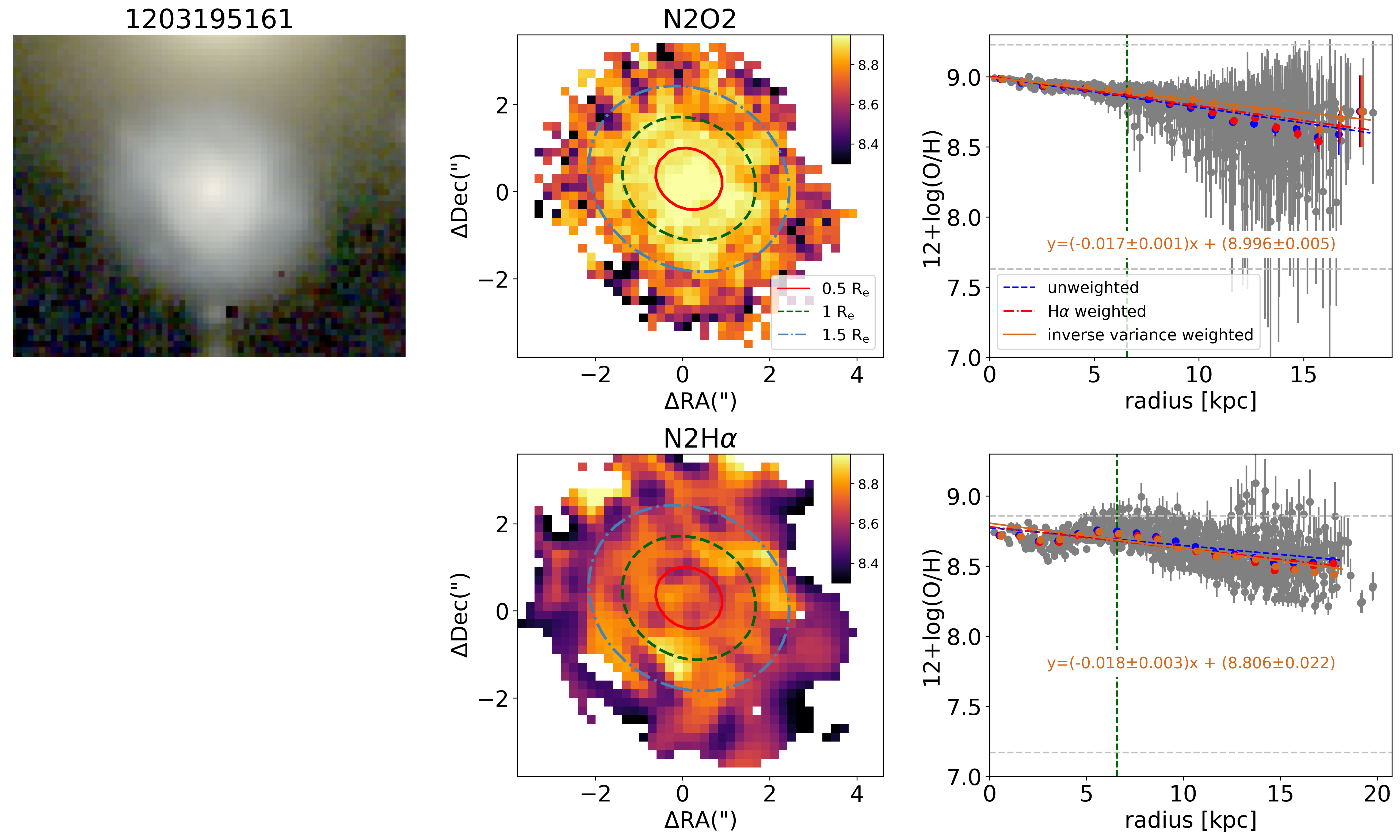}
    \caption{The metallicity and metallicity gradient measurement for \revise{an example} galaxy (MAGPIID 1203195161). \teamreview{The top-left panel shows a colour composite image of the galaxy, constructed from the $i$-, $r$-, and $g$-band data.} The first row shows the metallicity measured using the N2O2 diagnostic and the second row uses the N2H$\alpha$ diagnostic. The \teamreview{middle} panels show the seeing-deconvolved metallicity map. The right panels demonstrate the metallicity of each spaxel as a function of radius (grey) and the average metallicity in each 0.5 PSF$_\mathrm{FWHM}$ bin for three different weight methods. We overplot the linear best fits for each method. The horizontal lines in the right panels denote the valid metallicity range for each diagnostic \citep{Pettini:2004,Kewley:2019}. The green vertical lines denote 1 $R_\mathrm{e}$ for this galaxy. \reviewer{We show the value of the metallicity gradient and central metallicity (the inverse variance weighted method) for each plot.}}
    \label{fig:metal_example}
\end{figure*}

\begin{figure*}
    \centering
    \includegraphics[width=0.85\textwidth]{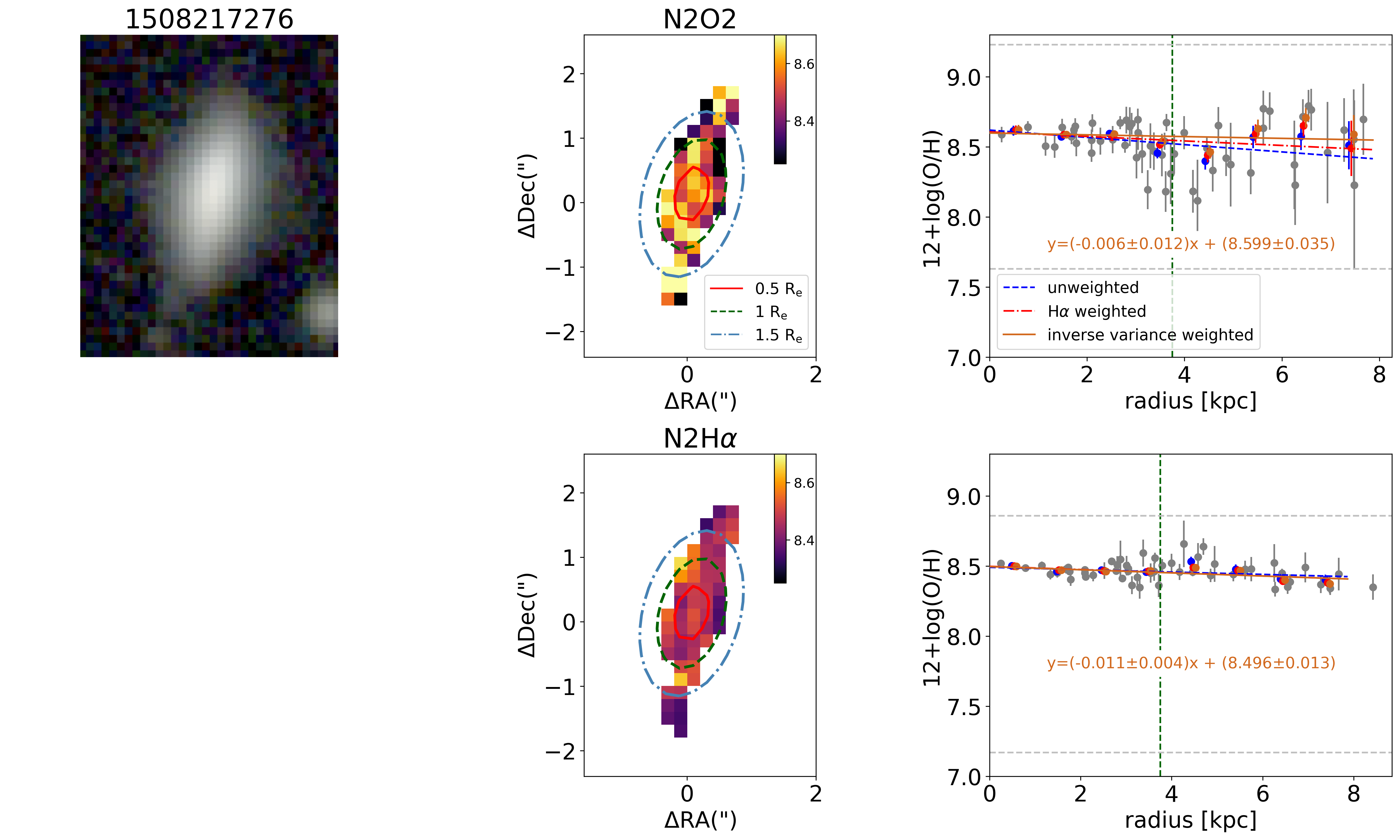}
    \caption{The metallicity and metallicity gradient measurement for an example galaxy (MAGPIID 1508217276). The same as Figure \ref{fig:metal_example}. }
    \label{fig:metal_example_typicalsize}
\end{figure*}

Figure \ref{fig:metal_example} and Figure \ref{fig:metal_example_typicalsize} show examples of the resolved metallicity map and metallicity gradient for two galaxies in the sample (MAGPIID 1203195161, MAGPIID 1508217276). \reviewer{The uncertainties were determined from the linear fitting process, performed using the \texttt{curve\_fit} function. This fitting routine accounts for the uncertainties in the average metallicity of individual bins, which were propagated through the fitting procedure to yield the uncertainty on the gradient.} \mnras{We also show the H$\alpha$, [N\,\textsc{ii}] and [O\,\textsc{ii}] flux distribution from the models made by \textsc{blobby3d} in Appendix \ref{sec:emission line flux}.}

\begin{figure*}
    \centering
    \includegraphics [width=0.75\textwidth]{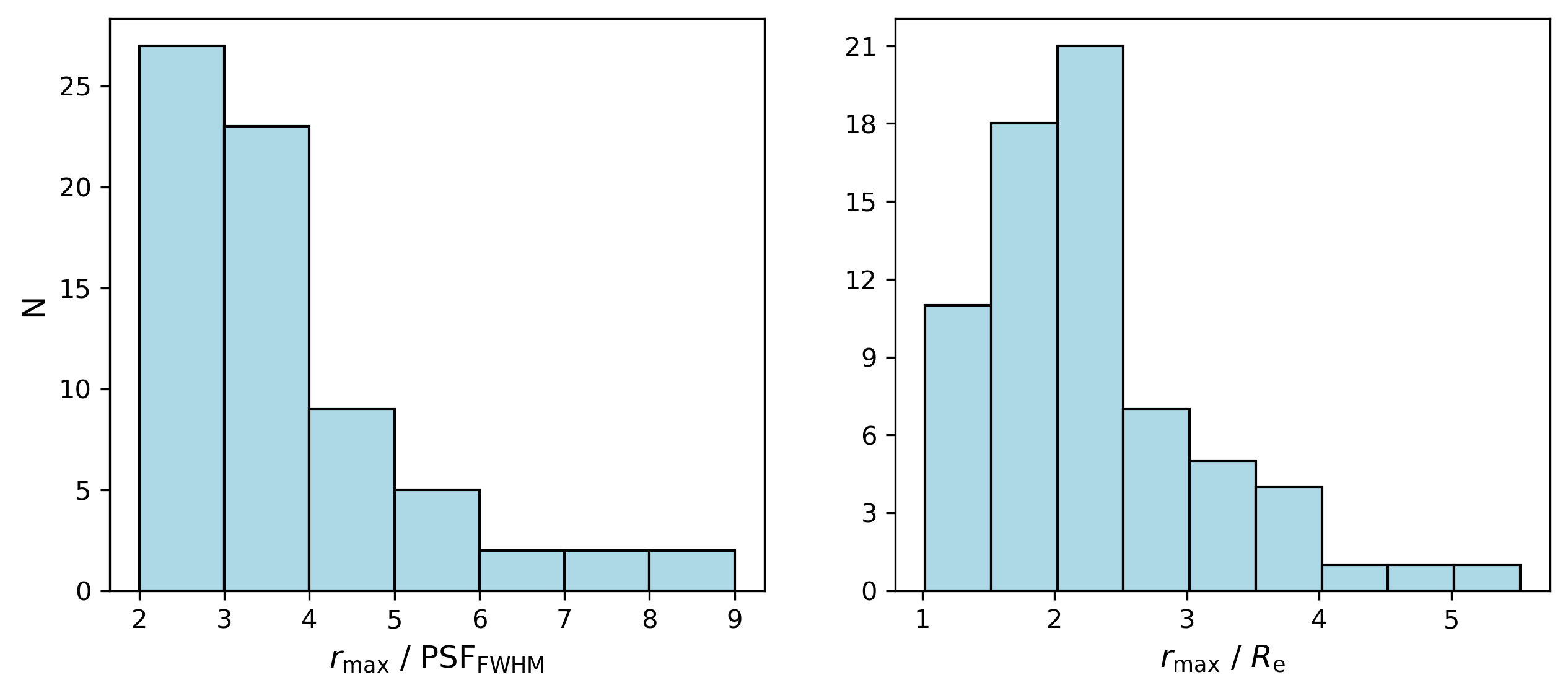}
    \caption{\mnras{The left panel shows the distribution of the radius of the most extended radial bin ($r_\mathrm{max}$) relative to PSF$_\mathrm{FWHM}$.} The right panel shows the distribution of $r_\mathrm{max}$ with respect to $R_\mathrm{e}$. \teamreview{Most of the galaxies are moderately resolved, extending out to a median of $R\sim2\,R_\mathrm{e}$}.}
    \label{fig:hist_binnumber}
\end{figure*}

\teamreview{We investigate how the spatial extent of galaxies affects the reliability of metallicity gradient measurements.} \paperrevise{Figure \ref{fig:hist_binnumber} shows \mnras{the distribution of the radius of the most extended radial bin ($r_\mathrm{max}$) relative to PSF$_\mathrm{FWHM}$ and to $R_\mathrm{e}$}. The majority (59/70) of our sample have gradients measured out to at least $1.5\,R_\mathrm{e}$. \mnras{In our sample selection, we include only those galaxies with gradients measured over a radial extent of at least 2\,PSF$_\mathrm{FWHM}$, as gradients derived from fewer resolution elements are subject to larger uncertainties.} The mean gradient uncertainty of the full sample is 0.01 $\mathrm{dex/kpc}$. \teamreview{To assess whether small galaxy size affects the accuracy of metallicity gradient measurements, we construct an extended subsample of galaxies with metallicity measured in at least 7 radial bins (0.5 PSF$_\mathrm{FWHM}$ each bin). For each of these galaxies, we re-fit the metallicity gradient using progressively smaller radial ranges, down to 4 bins.} We find good agreement between gradients of the same galaxies measured at small and large radii but with a typical offset of $\nabla \mathrm{[O/H]}[\ge7\, \mathrm{bins}]-\nabla \mathrm{[O/H]}[<7\, \mathrm{bins}]=0.02\, \mathrm{dex/kpc}$. We add this uncertainty in quadrature to the existing uncertainties for gradients measured with $<7$ bins.}

\section{Results and Analysis}\label{sec:results-4}

\reviewer{The median of the metallicity gradient\paperrevise{s} \teamreview{(N2O2 diagnostic, inverse variance weighted method)} is $-0.013^{+0.059}_{-0.033}$ dex/kpc. Among 70 galaxies in our sample, there are 23 galaxies (32.9\%) that have significantly negative metallicity gradients (i.e. deviations from zero greater than $2\sigma$), 7 galaxies (10.0\%) that have significantly positive metallicity gradients and 40 galaxies (57.1\%) that have flat gradients \revise{(i.e. consistent with zero within $2\sigma$)}.}

We use the metallicity measured with the N2O2 diagnostic and the inverse variance weighted method for the following analysis, unless otherwise specified. \teamreview{A comparison between the N2O2 and N2H$\alpha$ metallicity gradients is presented in Section \ref{sec: compare n2o2 and n2ha diagnostics}.} The metallicity gradients and central metallicities for all galaxies are compiled in Appendix \ref{sec:compiled data}, with results provided for both the N2O2 and N2H$\alpha$ diagnostics using the inverse variance weighted method. \teamreview{Using the inverse variance weighted method as our primary approach, we find that the metallicity gradients are consistent with those measured using unweighted and H$\alpha$ weighted methods. Specifically, 65 out of 70 galaxies are consistent within $0.5\,\sigma$, and all galaxies are consistent within $1\,\sigma$.}





\subsection{Correlation between metallicity gradient and galaxy properties}

To understand the mechanisms that influence metallicity gradients, we investigate the relation between metallicity gradient and several galaxy properties, including stellar mass ($M_*$), $R_\textup{e}$, a proxy for the gravitational potential ($M_*/R_\mathrm{e}$), $\Sigma_{M_*}$, SFR, $\Sigma_\textup{SFR}$, $\sigma_\textup{gas}$ and the \paperrevise{ratio between rotational velocity and gas velocity dispersion} ($v_{\phi}/\sigma_\mathrm{gas}$). 
\begin{figure*}
\centering
    \includegraphics [width=0.95\textwidth]{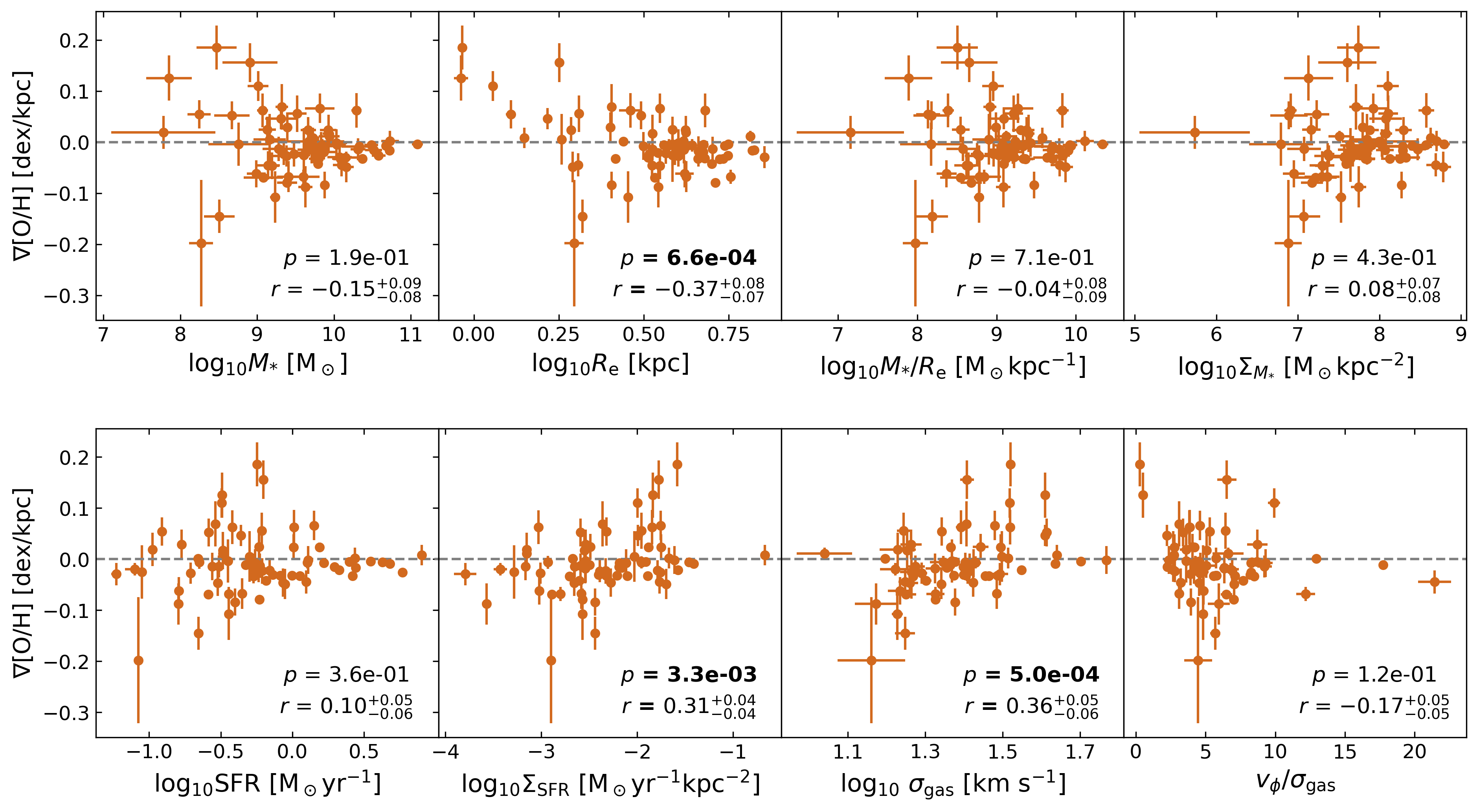}
    \includegraphics [width=0.95\textwidth]{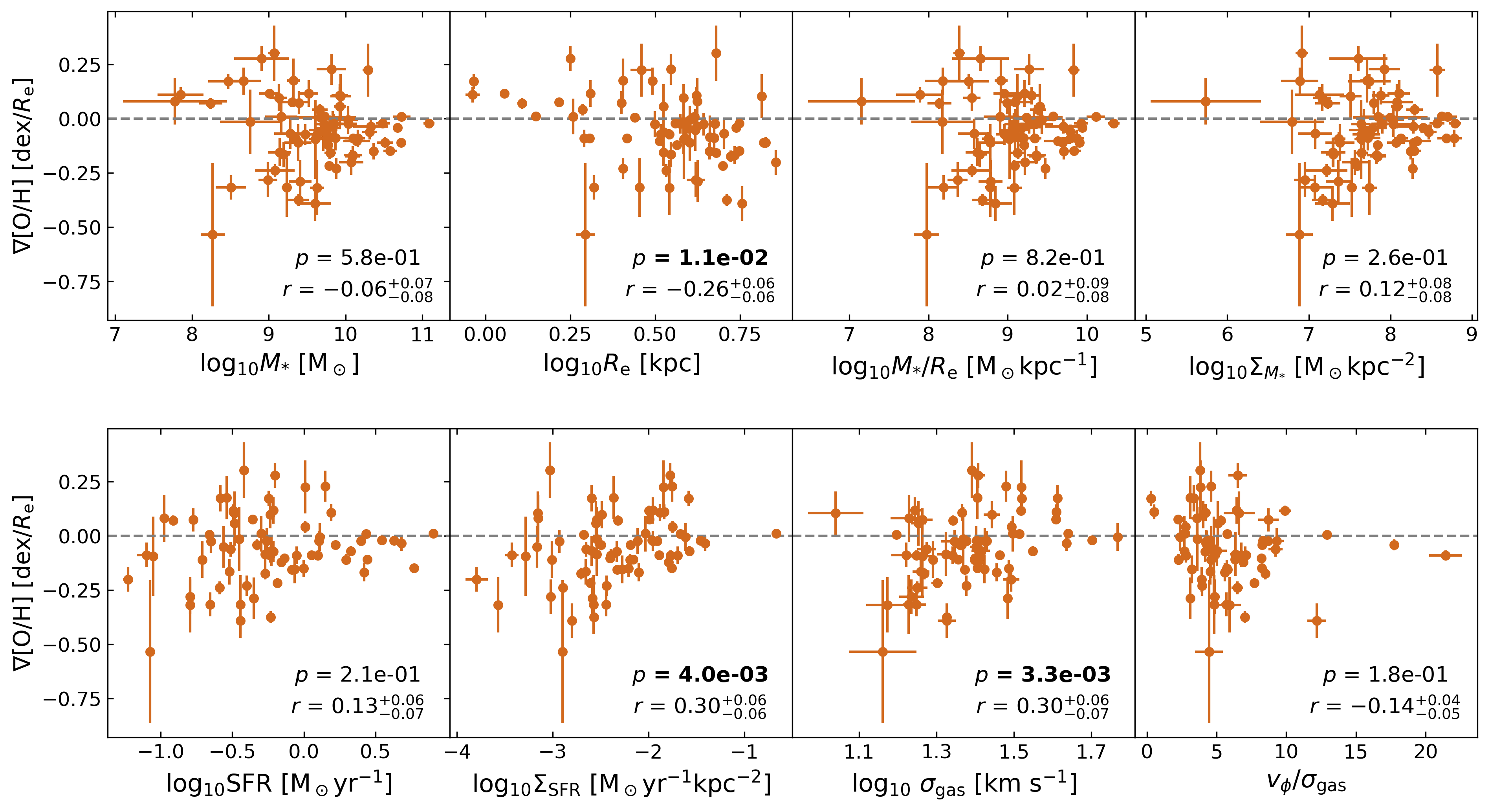}
    \caption{The N2O2 metallicity gradients (top two rows in units of dex/kpc, bottom two rows in units of dex/$R_\mathrm{e}$) as a function of a variety of galaxy properties, including stellar mass ($M_*$), $R_\textup{e}$, a proxy for the gravitational potential ($M_*/R_\mathrm{e}$), $\Sigma_{M_*}$, SFR, $\Sigma_\textup{SFR}$, $\sigma_\textup{gas}$ and $v_{\phi}/\sigma_\mathrm{gas}$. We show the Pearson correlation coefficient, $r$, and $p$-value on each panel. The horizontal dashed lines denote the flat gradient. \teamreview{The $\nabla \mathrm{[O/H]}$ of MAGPI galaxies has a moderately significant positive correlation with $\sigma_\textup{gas}$ and with $\Sigma_\textup{SFR}$ and negative correlation with $R_\mathrm{e}$.}}
    \label{fig:gradient_properties}
\end{figure*}
We show the relationship between the metallicity gradient ($\nabla \mathrm{[O/H]}$) and galaxy properties in Figure \ref{fig:gradient_properties}. 
\paperrevise{For each data point, we sample metallicity gradient and \mnras{galaxy property} values from a normal distribution centred on the measured value with a standard deviation equal to the uncertainty. We repeat this process 1000 times and calculate the Pearson correlation coefficient, $r$, for each resample. }
We show the median of $r$ and 16$^\mathrm{th}$ and 84$^\mathrm{th}$ percentiles of $r$ on each panel. We also show $p$-value, which helps determine whether the observed correlation is statistically significant. A $p$-value of 0.05 or lower is generally considered statistically significant. We show metallicity gradients in units of dex/kpc in the top two rows, and in units of dex/$R_\mathrm{e}$ in the bottom two rows.  The results are similar for metallicity gradients in different units. We discuss our results in units of dex/kpc.

\paperrevise{We find the $\nabla \mathrm{[O/H]}$ of MAGPI galaxies has a \teamreview{moderately} significant positive correlation with $\sigma_\textup{gas}$ ($r=0.36^{+0.05}_{-0.06},\ p=5.0\times  10^{-4}$) and with $\Sigma_\textup{SFR}$ ($r=0.31^{+0.04}_{-0.04},\ p=3.3\times  10^{-3}$). $R_\mathrm{e}$ is negatively correlated with the $\nabla \mathrm{[O/H]}$ of MAGPI galaxies ($r=-0.37^{+0.08}_{-0.07},\ p=6.6\times  10^{-4}$). There is no significant correlation between $\nabla \mathrm{[O/H]}$ and other parameters \teamreview{($|r|<0.2$ and $p > 0.05$)}.} 



\begin{figure}
    \centering
    \includegraphics [width=0.45\textwidth]{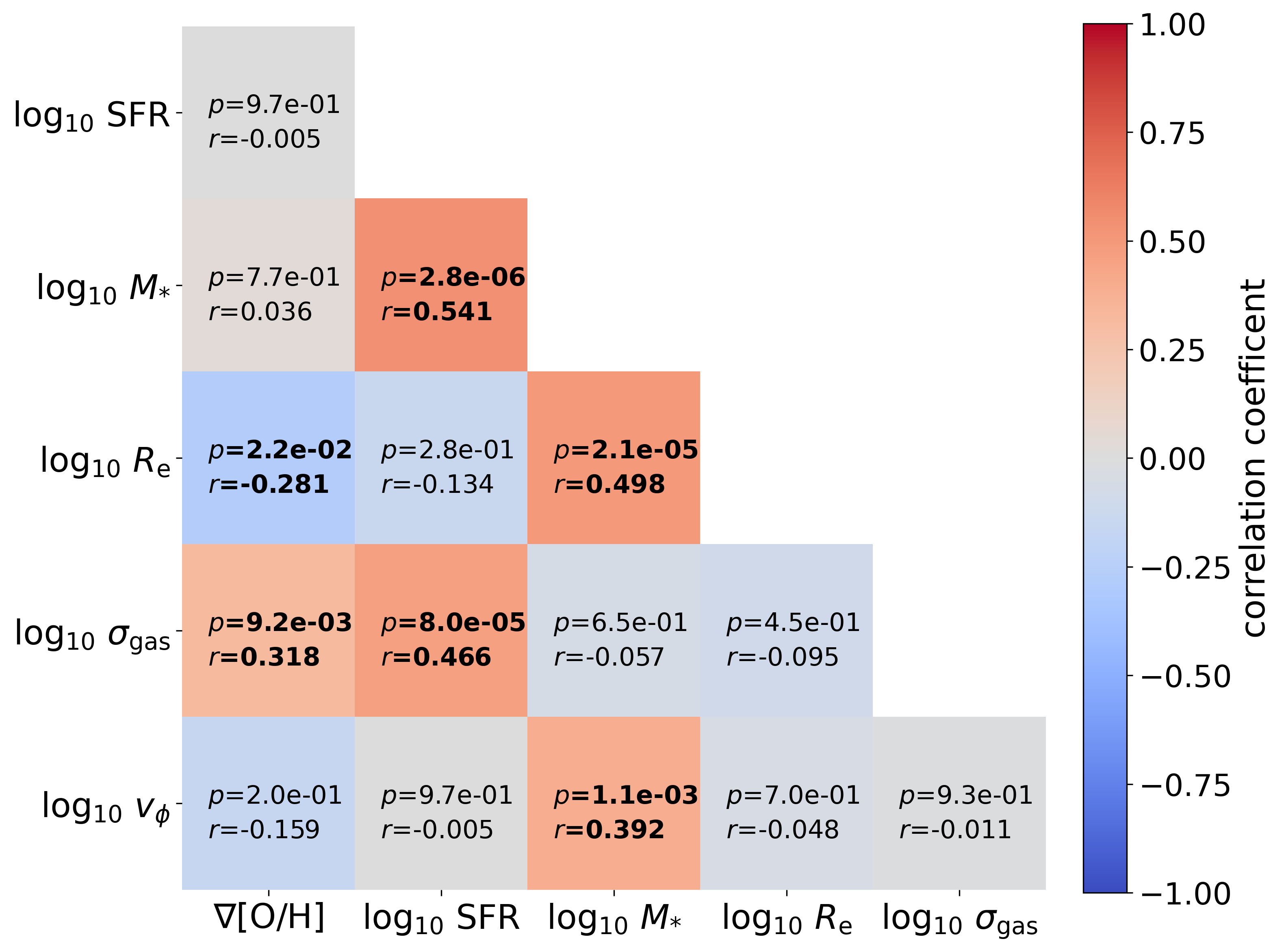}
    \caption{\mnras{The Pearson partial correlation between $\nabla \mathrm{[O/H]}$, SFR, $M_*$, $R_\textup{e}$, $\sigma_\mathrm{gas}$ and $v_\phi$ for MAGPI galaxies. The correlation coefficient ($r$) and $p$-value are shown in each box.}}
    \label{fig:partial_corr_metal}
\end{figure}

\reviewer{To disentangle the intercorrelation between different parameters, we also analyse the partial correlation \citep{Whittaker:2009,Kim:2015} between $\nabla \mathrm{[O/H]}$, SFR, $M_*$, $R_\textup{e}$, $\sigma_\mathrm{gas}$ and $v_\phi$. We calculate the Pearson partial correlation coefficient for each pair of parameters and $p$-value using \textsc{Pingouin} \citep{Vallat2018}. \mnras{The partial correlation quantifies the linear relationship between two variables while statistically controlling for the influence of all other variables in the sample. This approach isolates the direct association between each pair of parameters, independent of any shared correlations with the remaining variables.} We consider the correlation to be significant when $p<0.05$. \paperrevise{As shown in Figure \ref{fig:partial_corr_metal}, $\nabla \mathrm{[O/H]}$ has the strongest partial correlation with $\sigma_\mathrm{gas}$. In addition to $\sigma_\mathrm{gas}$, $R_\textup{e}$ has a negative correlation with $\nabla \mathrm{[O/H]}$. The correlation between $\nabla \mathrm{[O/H]}$ and $M_*$, SFR and $v_\phi$ is not significant \teamreview{when accounting for the correlation with $\sigma_\mathrm{gas}$}.}}

\subsection{Comparison of the N2O2 and N2H$\alpha$ diagnostics} \label{sec: compare n2o2 and n2ha diagnostics}

\begin{figure}
\centering
    \includegraphics [width=0.45\textwidth]{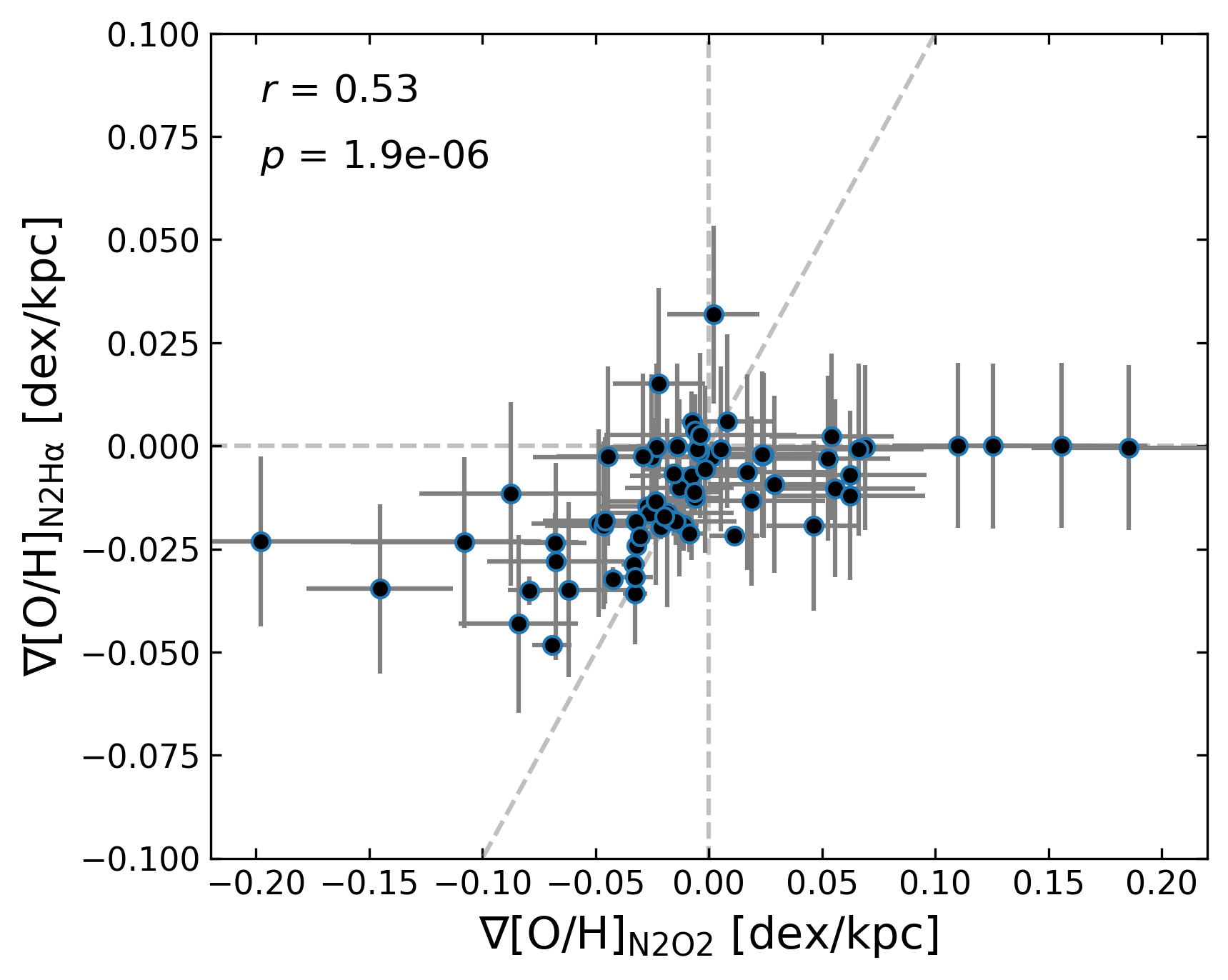}
    \caption{\revise{The comparison of the metallicity gradients measured using N2O2 and N2H$\alpha$ diagnostics. The diagonal dashed line indicates the one-to-one relation. The correlation coefficient ($r$) and $p$-value are shown in the top left.} \teamreview{The metallicity gradients measured using two diagnostics are significantly correlated ($r = 0.53$, $p = 1.9 \times 10^{-6}$), though data show substantial scatter around the one-to-one line.}}
    \label{fig:n2o2_n2ha_comapre}
\end{figure}

We compare the N2O2 and N2H$\alpha$ metallicity gradients in Figure \ref{fig:n2o2_n2ha_comapre}. \teamreview{While the metallicity gradients measured using these two diagnostics are significantly correlated ($r=0.53,\ p=1.9\times  10^{-6}$), they exhibit substantial scatter and deviate from the one-to-one line.} These differences may arise from the differing sensitivity of the two diagnostics to variations in ionisation parameter and ISM pressure \citep[e.g.][]{Poetrodjojo:2021}.

Our results show that different metallicity diagnostics predict different metallicity distributions in some galaxies. For example, in galaxy MAGPIID 1203195161 (Figure \ref{fig:metal_example}) the metallicity decreases with radius smoothly in the N2O2 diagnostic, while in the N2H$\alpha$ diagnostic, the metallicity decreases within \reviewer{4 kpc} and then slightly increases between \reviewer{$4-7$ kpc} and decreases again at the larger radii. The variation of ionisation parameter and ISM pressure across the galaxy may play a role in the increase of the metallicity profile, as the N2H$\alpha$ diagnostic is sensitive to ionisation parameter \teamreview{and sensitive to ISM pressure at high metallicities}. Therefore, assuming a constant log($U$) limits the accuracy of the N2H$\alpha$ diagnostic. This finding aligns with \citet{Poetrodjojo:2021}, who showed that the metallicity gradients derived using different diagnostics present systematic discrepancies. \citet{Poetrodjojo:2021} reported a Spearman rank correlation coefficient of $\rho=0.59$ between the N2O2 and N2H$\alpha$ metallicity gradients, which is comparable in strength to the Pearson correlation we measure ($r=0.53$).

%
\teamreview{Our results are comparable to the findings of \citet{Easeman:2022}, who explored the origin of dips in radial metallicity profiles using various diagnostics. While they did not find a clear link between these dips and the variation of ionisation parameter, they proposed that inside-out quenching could be a possible driver. }\reviewer{In our sample, some galaxies have elevated N2H$\alpha$ metallicity in their centres or their outer discs. This may link to inside-out quenching or outside-in quenching, which depends on the environment \citep[e.g.][]{Wang:2022}. Further studies are needed to investigate how the variation of ionisation parameter and quenching scenarios influences the metallicity measured using different diagnostics.}

\teamreview{While the strength of the correlations differs, the trends in metallicity gradients derived from N2H$\alpha$ and N2O2 are broadly consistent in direction. The N2H$\alpha$ metallicity gradient in the unit of dex/kpc shows a weak positive correlation with $\sigma_\mathrm{gas}$ ($r=0.20^{+0.09}_{-0.10},\ p=9.3\times  10^{-3}$) and $\Sigma_\mathrm{SFR}$ ($r=0.22^{+0.08}_{-0.09},\ p=4.5\times  10^{-3}$). Compared with the N2O2 metallicity gradient, the N2H$\alpha$ gradient has no significant correlation with $R_\mathrm{e}$ ($r=-0.12^{+0.11}_{-0.10},\ p=8.5\times  10^{-2}$).} 

\begin{figure}
\centering
    \includegraphics [width=0.45\textwidth]{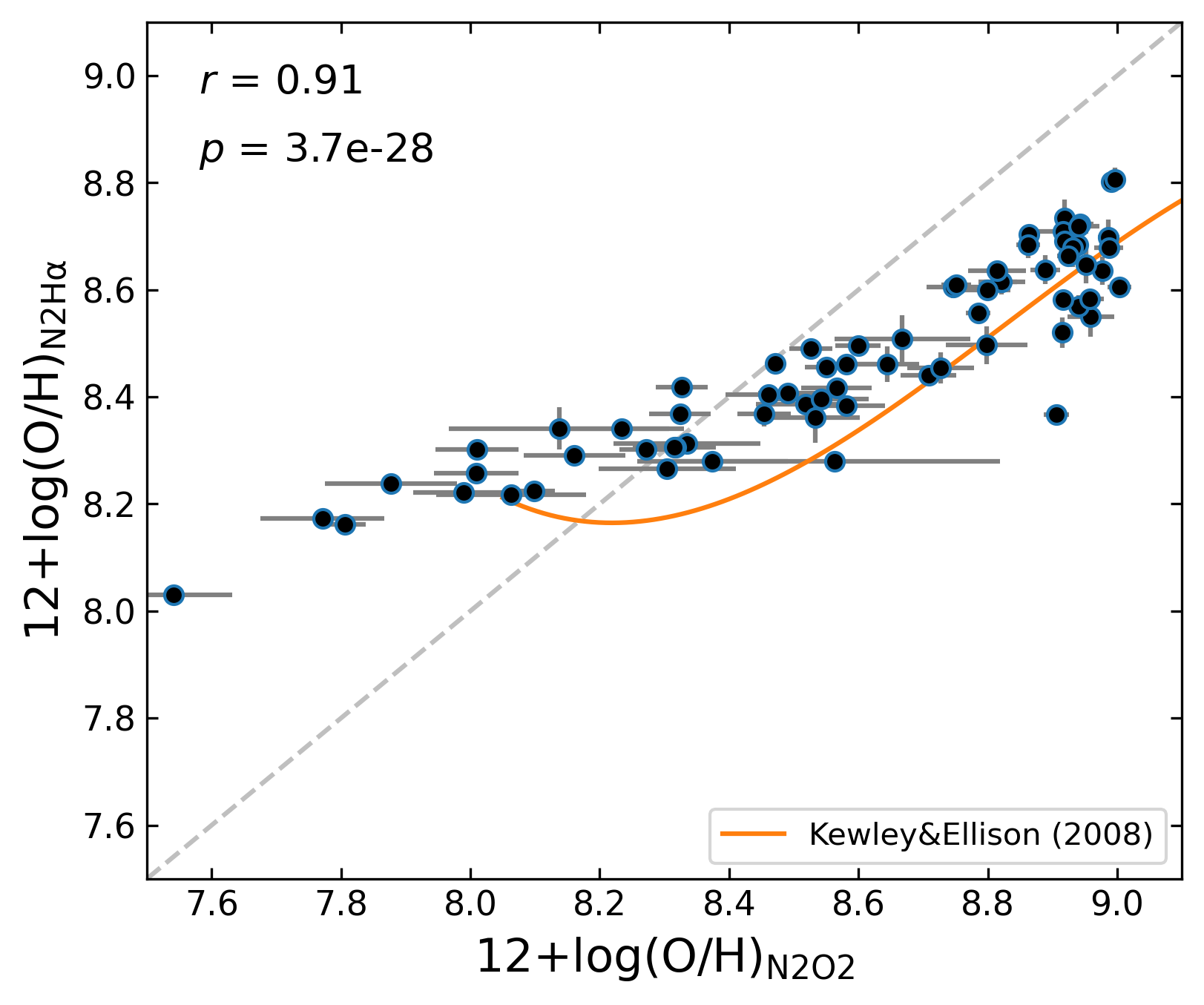}
    \caption{\mnras{The comparison of the central metallicity measured using N2O2 and N2H$\alpha$ diagnostics. The diagonal dashed line indicates the one-to-one relation. The correlation coefficient ($r$) and $p$-value are shown in the top left. The solid line shows the empirical calibration conversion between N2O2 and N2H$\alpha$ from \citet{Kewley:2008}. The two diagnostics are tightly correlated ($r=0.91$, $p=3.7\times10^{-28}$), with N2O2 yielding systematically higher metallicities by $\sim$0.2 dex at the high-metallicity end and slightly lower values at the low-metallicity end. The trend is consistent with the empirical calibration conversion by \citet{Kewley:2008}.}}
    \label{fig:n2o2_n2ha_comapre_central}
\end{figure}

\mnras{We show the comparison of the central metallicity derived from the two calibrations in Figure \ref{fig:n2o2_n2ha_comapre_central}. The central metallicities measured using the N2O2 and N2H$\alpha$ diagnostics are tightly correlated ($r = 0.91$, $p = 3.7\times10^{-28}$). The N2O2 diagnostic yields $\sim0.2$ dex higher metallicities than N2H$\alpha$ at the high-metallicity end, while the two diagnostics give consistent results around $12+\log(\mathrm{O/H})\sim8.3$. At the low-metallicity end, the N2O2 diagnostic predicts slightly lower metallicities. Overall, the relation between the N2O2- and N2H$\alpha$-based metallicities is consistent with the empirical calibration conversion determined by \citet{Kewley:2008}.}

\subsection{The influence of seeing on MAGPI metallicity gradient measurements} \label{sec:seeing effect}



\begin{figure*}
    \centering
    \includegraphics [width=0.75\textwidth]{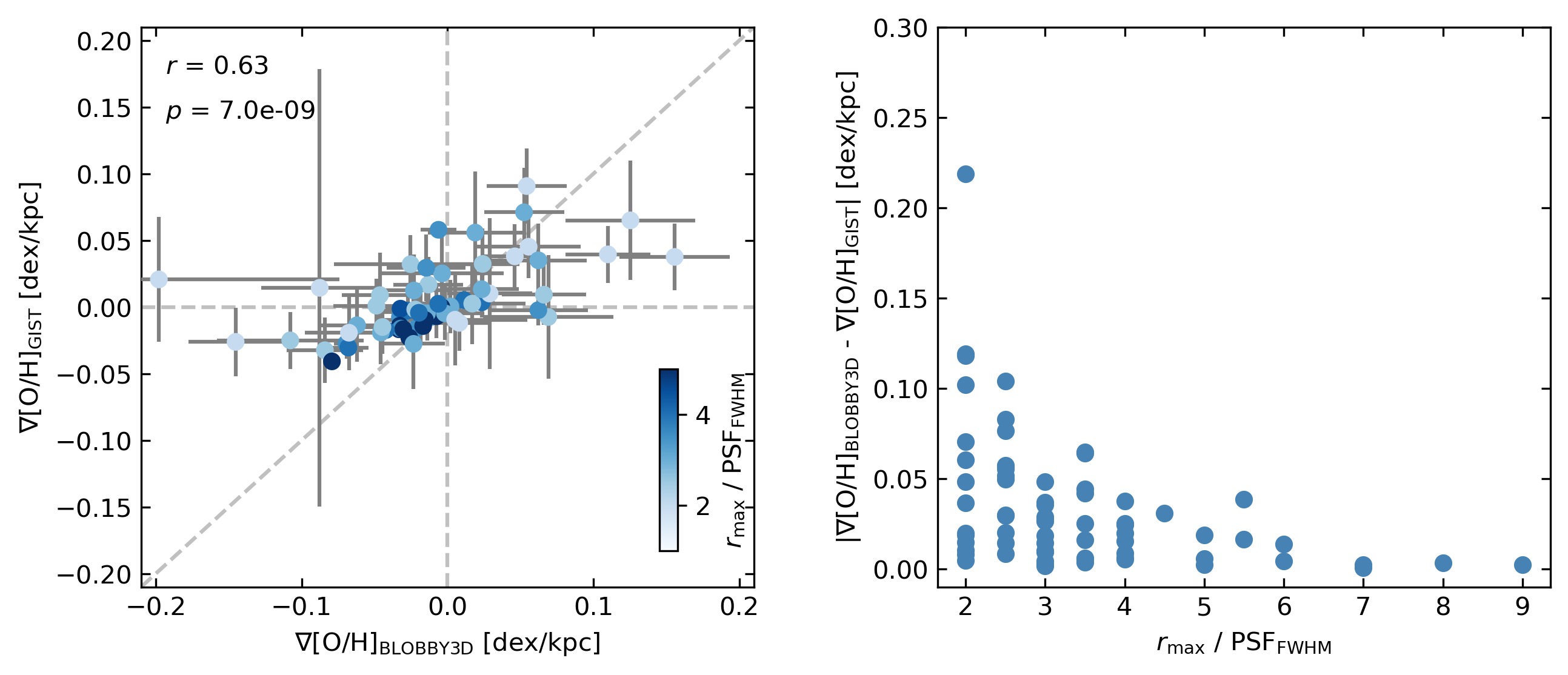}
    \caption{Left: Comparison of seeing-convolved ($\nabla \mathrm{[O/H]}_\mathrm{GIST}$) and seeing-deconvolved ($\nabla \mathrm{[O/H]}_\mathrm{BLOBBY3D}$) metallicity gradients of the MAGPI galaxies, colour-coded by the ratio between the radius of the maximum radial bin ($r_{\mathrm{max}}$) and the FWHM of PSF. \mnras{The correlation coefficient ($r$) and $p$-value are shown in the top left.} \paperrevise{Right: The difference between $\nabla \mathrm{[O/H]}_\mathrm{GIST}$ and $\nabla \mathrm{[O/H]}_\mathrm{BLOBBY3D}$ versus  $r_\mathrm{max}$/PSF$_\mathrm{FWHM}$.}}
    \label{fig:gist_b3d_compare}
\end{figure*}

The modelling technique we used accounts for the effect of seeing and has a large flexibility \revise{in} the flux distribution. We compare the seeing-deconvolved metallicity gradient measured from the flux map output of \textsc{blobby3d} with the seeing-convolved metallicity gradient measured from the flux map output of \textsc{gist}, in Figure \ref{fig:gist_b3d_compare}. \paperrevise{As expected,} we see that the seeing-convolved metallicity gradients are flatter than the seeing-deconvolved gradients \citep{Yuan:2013,Stott:2014,Belfiore:2017,Wuyts:2016,Acharyya:2020}. \paperrevise{The seeing-convolved gradients are closer to seeing-deconvolved gradients when the galaxies are large compared to the seeing. The difference is negligible ($\Delta(\nabla \mathrm{[O/H]}) <0.014$ dex/kpc) for galaxies with gradients measured beyond  $r_\mathrm{max} / \mathrm{PSF}_\mathrm{FWHM} \geqslant 6$, as shown in the right panel of Figure \ref{fig:gist_b3d_compare}. } \teamreview{For well-resolved galaxies, the difference between seeing-convolved and deconvolved gradients is minor. For more compact galaxies, however, the impact of seeing could be significant, making it essential to account for PSF effects in studies of metallicity gradients.}


\section{Discussion}\label{sec:discussion-4}

\subsection{Drivers of the metallicity gradient}

The MAGPI galaxies in our sample show mildly negative metallicity gradients on average, a trend that is generally interpreted as a consequence of disc inside-out growth, where the central regions experience longer star formation and enrichment than the outskirts \citep{Larson:1976,Matteucci:1989,Boissier:1999}. However, a substantial fraction of galaxies display flat gradients, and a smaller fraction even positive gradients, pointing to the influence of additional processes.


Galaxies can have a steep negative metallicity gradient if the metals stay undisturbed in the galactic disc \citep{Gibson:2013}. However, mechanisms like gas inflows, outflows, accretion, galaxy interaction \teamreview{and mergers} play a role in redistributing the metals in the galaxy, which can flatten or even invert the metallicity gradient \citep{Cresci:2010,Kewley:2010,Rupke:2010,Marinacci:2014,Chisholm:2018}. 
\reviewer{The positive correlation between $\nabla \mathrm{[O/H]}$ and $\sigma_\mathrm{gas}$ of the MAGPI galaxies (Figure \ref{fig:gradient_properties}) suggests a connection between gas turbulence and metallicity gradient flattening. \citet{Mai:2024} found that the main driver of gas turbulence in the MAGPI galaxies is stellar feedback or gas instability. \paperrevise{They found that $\sigma_\mathrm{gas}$ has \teamreview{the strongest correlation with $\Sigma_{\rm SFR}$ and} the second strongest correlation with the non-rotational motion of gas \teamreview{(e.g. gas bulk inflows and outflows)}, suggesting that} gas accretion and transportation also contribute to gas turbulence in the MAGPI galaxies. This result agrees with the model for metallicity gradients developed by \citet{Sharda:2024}, which showed that stellar feedback, gas accretion and transportation are important to drive gas turbulence and metal-mixing. Only models that include those mechanisms can reproduce the $\nabla \mathrm{[O/H]}-M_*$ relation in the local galaxies.}

\teamreview{The connection between gas velocity dispersion and metallicity gradients has been reported in high-redshift galaxies. \citet{Queyrel:2012} found a weak positive correlation ($r = 0.26$) in a sample of 26 galaxies at $z \sim 1.2$. In comparison, our sample at $z \sim 0.3$ yields a stronger correlation ($r = 0.36$) based on 70 galaxies. Similarly, \citet{Sharda:2021b} compiled a sample of 74 high-redshift galaxies ($0.1 \leq z \leq 2.5$) from multiple ground-based IFU measurements and found that most of the dispersion-dominated galaxies ($v_\phi / \sigma_\mathrm{gas} \leq 1$, where $v_\phi$ is the rotational velocity) have flat gradients, while the rotation-dominated galaxies have both flat gradients and steep negative gradients. However, we find the relation between $\nabla \mathrm{[O/H]}$ and $v_\phi / \sigma_\mathrm{gas}$ is insignificant in the MAGPI sample.}

Our result also shows that galaxies with higher $\Sigma_\mathrm{SFR}$ have flatter \paperrevise{or positive} metallicity gradients (Figure \ref{fig:gradient_properties}). \revise{This} trend is consistent with the scenario that the accretion of pristine gas or gas transportation from the outskirts to the inner regions flattens \paperrevise{or inverts} the metallicity gradient and triggers star formation \citep{Rupke:2010,Stott:2014}. 
\reviewer{The galaxies with positive metallicity gradients} might result from metal-poor intergalactic medium \revise{accreting} directly to the centre of the galaxy, which dilutes the central metallicity and inverts the metallicity gradient \citep{Cresci:2010}. \teamreview{High values of $\Sigma_\mathrm{SFR}$ are also linked to a high likelihood of the presence of star-formation driven winds \citep[e.g.][]{Martin:2005,Rupke:2005,Heckman:2016}, which can dilute gradients.} \mnras{Partial correlation analysis (Figure \ref{fig:partial_corr_metal}) shows that the $\nabla \mathrm{[O/H]}$–$\Sigma_\mathrm{SFR}$ relation is largely driven by the underlying dependence on galaxy size, with compact galaxies (small $R_\mathrm{e}$) more likely to exhibit flat or positive gradients. This suggests that the mechanisms that redistribute metals are likely more efficient in smaller galaxies.}

\begin{figure*}
    \centering
    \includegraphics [width=0.8\textwidth]{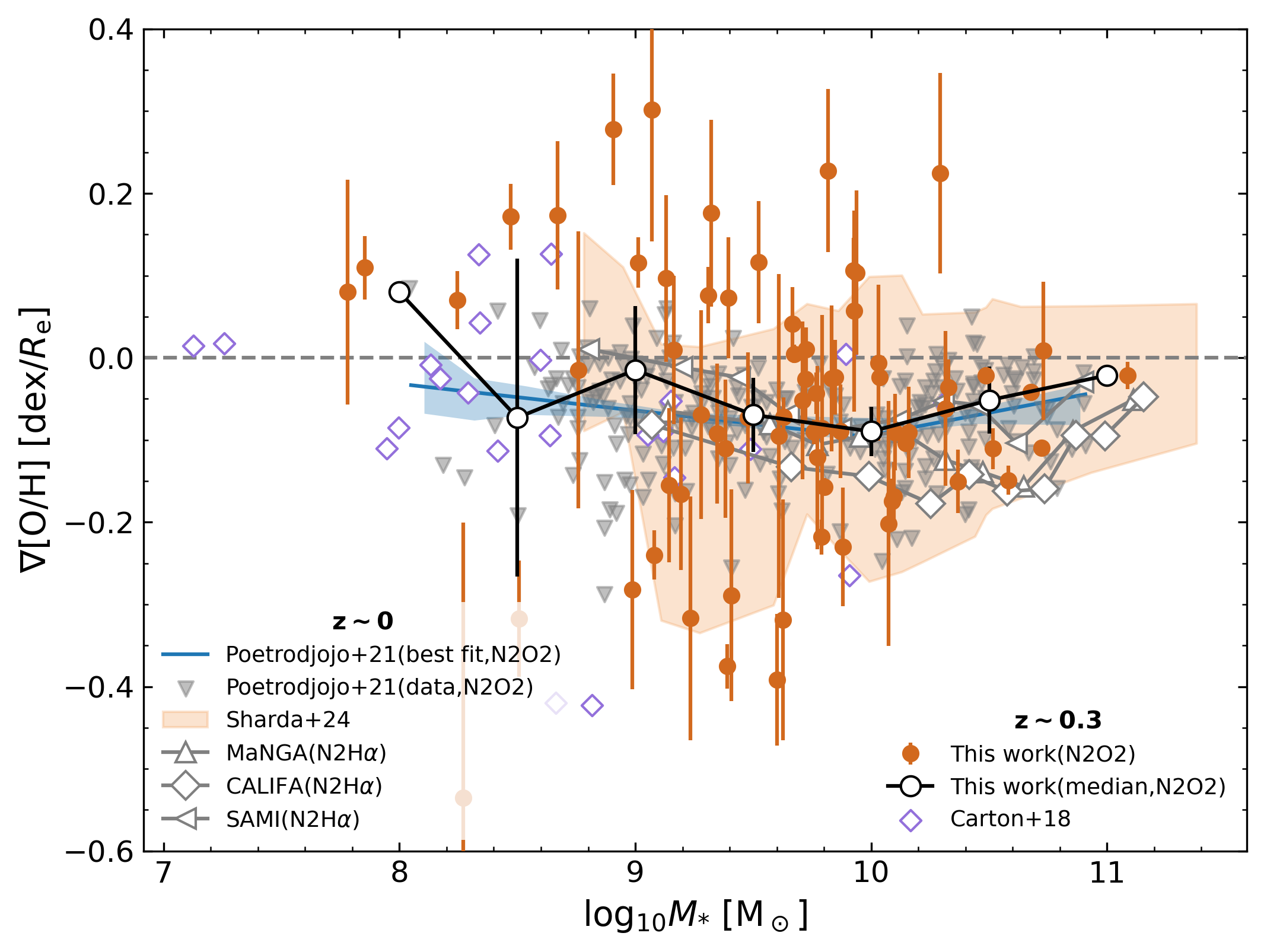}
    \caption{\reviewer{The $\nabla \mathrm{[O/H]}-M_*$ relation for galaxies at $z\sim 0.3$ and $z\sim 0$. The galaxies of the MAGPI survey are shown as solid orange circles. The median $\nabla \mathrm{[O/H]}$ in each 0.3 dex $M_*$ bin are shown as black edge circles. The galaxies at the same redshift range as this study ($0.26 < z < 0.42$) measured by \citet{Carton:2018} are shown as open purple diamonds. The galaxies of the SAMI survey at $z \sim 0$ are shown as grey triangles (N2O2 diagnostic) and their best fit (a broken linear-fit) is shown as a blue line and the 1$\sigma$ uncertainty is shown as a blue band. \reviewer{The model prediction from \citet{Sharda:2024} for the galaxies in the local Universe is shown as a pink band.} The galaxies in the local Universe from the SAMI \citep{Poetrodjojo:2021}, MaNGA \citep{Belfiore:2017} and CALIFA \citep{Sanchez:2014} surveys are shown as grey-edge markers, with seeing-correction made by \citet{Acharyya:2022}. Note that the metallicity of galaxies in the SAMI, MaNGA, and CALIFA surveys shown as grey-edge markers is measured using the N2H$\alpha$ diagnostic.}}
    \label{fig:gradient_mass}
\end{figure*}

The relative strength of different mechanisms may vary with stellar mass. We find a bend in the $\nabla \mathrm{[O/H]}-M_*$ relation in our sample, with the minimum around $10^{10}\ \mathrm{M}_\odot$, as shown in Figure \ref{fig:gradient_mass}. Observations of local galaxies suggest that the mass-metallicity gradient relation reaches a minimum at $\sim 10^{10}\ \mathrm{M}_\odot$ for N2O2-based gradients (or $\sim 10^{10.5}\ \mathrm{M}_\odot$ for N2H$\alpha$ gradients), i.e. the metallicity gradient decreases with increasing stellar mass when $M_* \lesssim 10^{10}\ \mathrm{M}_\odot$ and increases with increasing stellar mass when $M_* \gtrsim 10^{10}\ \mathrm{M}_\odot$ \citep{Belfiore:2017,Mingozzi:2020,Poetrodjojo:2021,Li:2025}. \teamreview{The existence of the bend is insensitive to the choice of the metallicity diagnostic \citep{Poetrodjojo:2021}. Theoretical models presented in \citet{Sharda:2024} \mnras{ explored the relative contribution of stellar feedback, gas accretion, transportation and winds in shaping metallicity profiles.} They proposed the following physical interpretation for this behaviour in local galaxies: (1) In low-mass galaxies ($M_* \lesssim 10^{9.5}\ \mathrm{M}_\odot$), metal mixing processes, such as gas accretion and feedback-driven outflows, dominate over in-situ metal production, resulting in relatively flat or even inverted metallicity gradients. (2) In intermediate-mass galaxies ($10^{9.5} \lesssim M_* \lesssim 10^{10.5}\ \mathrm{M}_\odot$), metal production becomes more efficient, while radial gas flows are weaker, leading to steeper metallicity gradients. (3) In high-mass galaxies ($ M_* \gtrsim 10^{10.5}\ \mathrm{M}_\odot$), gas accretion and transportation again become dominant over metal production, leading to the flattening of metallicity gradients at the high-mass end.}


We compare the $\nabla \mathrm{[O/H]}-M_*$ relation of the MAGPI galaxies \reviewer{with} galaxies at the same redshift\footnote{\revise{Note that the redshift range of the galaxies in \citet{Carton:2018} is $0.1 \lesssim z \lesssim 0.8$. We only select galaxies from their sample that are in the same redshift range as this study ($0.26<z<0.42$).}} \citep[$z\sim 0.3$;][]{Carton:2018}, galaxies at $z\sim 0$, and the theoretical predictions \reviewer{for local galaxies from \citet{Sharda:2024}, with the mass-loading factor in the model estimated from the EAGLE cosmological simulations \citep{Mitchell:2020}, as shown in Figure \ref{fig:gradient_mass}.}  
The $\nabla \mathrm{[O/H]}-M_*$ relation of the MAGPI galaxies qualitatively agrees with the prediction by \citet{Sharda:2024}. \paperrevise{The MAGPI galaxies have similar metallicity gradients on average to the galaxies in the SAMI, MaNGA and CALIFA samples (all at $z\sim 0$). However, the scatter and uncertainty of the metallicity gradient are large in our low-mass bins ($M_* \leq 10^{9.5}\ \mathrm{M}_\odot$). The model in \citet{Sharda:2024} shows that the metallicity gradient of low-mass galaxies ($M_* \lesssim 10^{9.5}\ \mathrm{M}_\odot$) varies significantly with the extent of newly produced metals mixing with the interstellar medium \revise{(described by the yield reduction factor, $\phi_y$\footnote{\reviewer{$0\leq \phi_y \leq 1$. $\phi_y \sim 1$ means the newly produced metals from supernovae are fully mixed with ISM before ejected and $\phi_y \sim 0$ means the newly produced metals are ejected from the galaxy without mixing with ISM.}}).} The low-mass bins in our sample are dominated by positive or flat metallicity gradients, consistent with local dwarf samples \citep{Li:2025}. Such positive gradients may arise from the accretion of metal-poor gas \citep{Cresci:2010} \teamreview{or feedback-driven outflows}, as has been suggested for dwarf galaxies \citep{Grossi:2020,Li:2025}. A larger sample, or higher spatial resolution, is needed to confirm this result.}




\subsection{Evolution of metallicity gradient}

\begin{figure*}
    \centering
    \includegraphics [width=\textwidth]{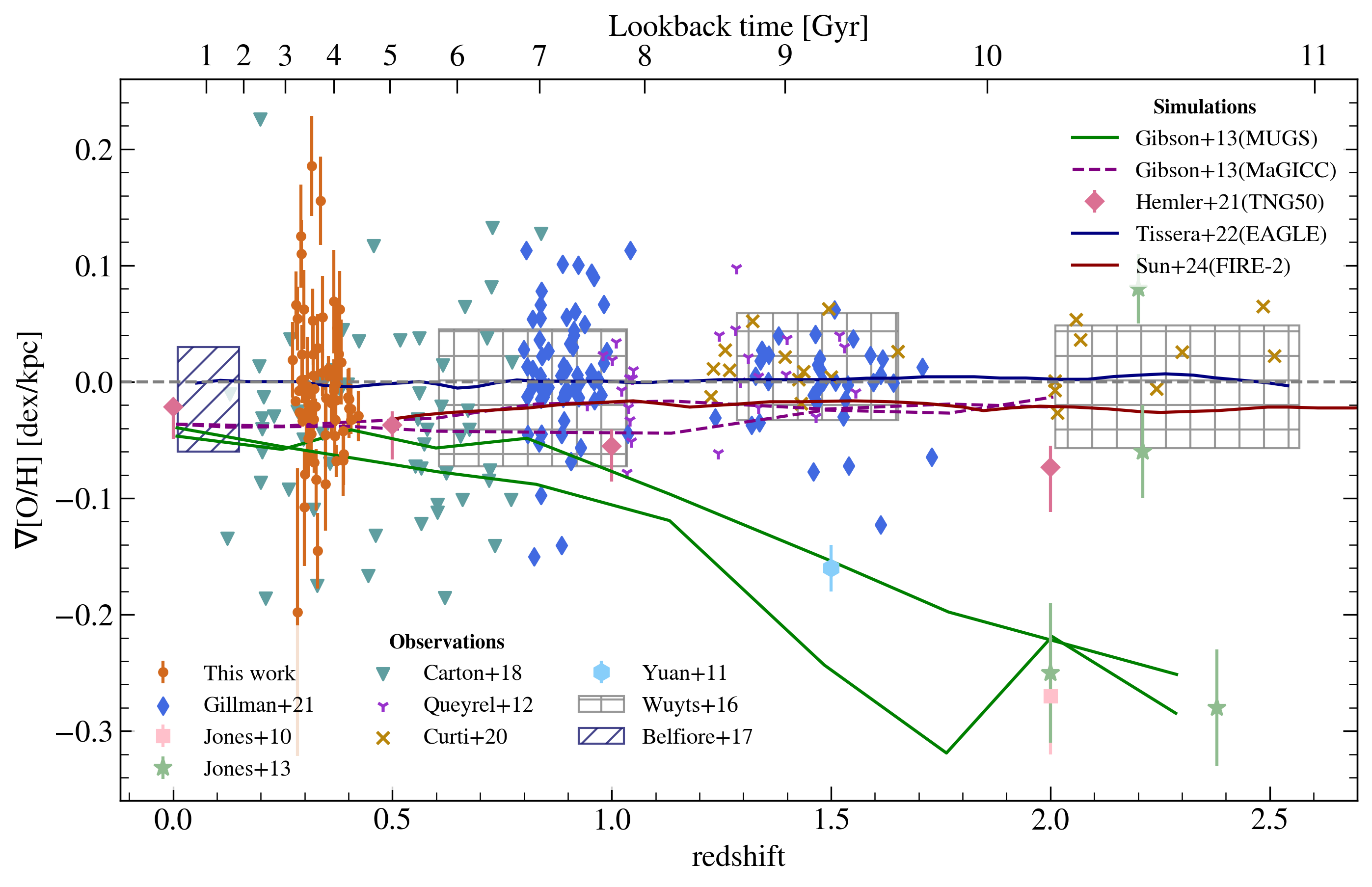}
    \caption{The redshift evolution of the metallicity gradient from previous studies and this work. We compile simulation results from \citet{Gibson:2013}, \citet{Helmer:2021}, \citet{Tissera:2022} and \citet{Sun:2024}. We also include observational studies results from \citet{Gillman:2021}, \citet{Carton:2018}, \citet{Wuyts:2016}, \citet{Belfiore:2017}, \citet{Jones:2010,Jones:2013}, \citet{Yuan:2011} \reviewer{, \citet{Queyrel:2012} and \citet{Curti:2020}}. We note that the regions with diagonal and cross-hatching show the range of metallicity gradient and redshift in \citet{Belfiore:2017} and \citet{Wuyts:2016}, respectively. \teamreview{The metallicity gradient shows only mild evolution with redshift, while a large intrinsic scatter is present at all epochs.}}
    \label{fig:gradient_redshift}
\end{figure*}

\paperrevise{Most of the observational and simulation studies show that the evolution of metallicity gradients is mild with redshift \citep[e.g.][]{Wuyts:2016,Gillman:2021,Helmer:2021,Tissera:2022,Sun:2024}. 
However, there is a higher fraction of galaxies displaying positive metallicity gradients at $z\gtrsim1$ than at $z \sim 0$ \citep{Cresci:2010,Wuyts:2016,Carton:2018,Gillman:2021}.} 
We compile metallicity gradient measurements from previous studies, in combination with our results, in Figure \ref{fig:gradient_redshift}. The metallicity gradients of the MAGPI galaxies generally agree with previous observations \citep{Carton:2018} and simulations \citep{Gibson:2013,Tissera:2022,Sun:2024} results at the same redshift. 

 \paperrevise{Among all observational studies presented in Figure \ref{fig:gradient_redshift}, only \citet{Carton:2018} and this work employ the forward modelling approach to account for beam smearing. In this study, we model the flux distributions of individual emission lines and subsequently derive the metallicity gradient, whereas \citet{Carton:2018}  directly modelled the metallicity gradient. Both \citet{Wuyts:2016} and \citet{Gillman:2021} modelled the impact of beam smearing as a function of different observational and intrinsic parameters and applied corrections to each galaxy based on their model.} \paperrevise{The scatter of the metallicity gradients of our result is similar to the results in \citet{Carton:2018}, but notably larger than the seeing convolved results. We also note that we are able to measure metallicity gradients down to stellar masses much lower ($<10^{9}\ \rm{M\odot}$) than those of the studies at $z\gtrsim1$. }

The divergence of the metallicity gradient predictions is obvious for galaxies at $z \gtrsim 1$. \citet{Gibson:2013} demonstrated the evolution of the metallicity gradient of two types of feedback models, as shown in Figure \ref{fig:gradient_redshift}. MaGICC \revise{\citep[Making Galaxies in a Cosmological Context;][]{Brook:2012b}} is the `enhanced' feedback model, which is efficient in redistributing the metals over large scales and flattening the metallicity gradient at high redshift. MUGS \revise{\citep[McMaster Unbiased Galaxy Simulations;][]{Stinson:2010}} is the `conventional' feedback model, which employed a weaker feedback scheme than MaGICC, returning much steeper metallicity gradients for galaxies at high redshift. \reviewer{We note that the stellar mass range of the galaxies in MaGICC and MUGS simulation is comparable to the MAGPI data we use.}

\teamreview{It is unclear which of these two feedback models best describes the observed metallicity gradients at $z \sim 0.3$.} \paperrevise{However, our result shows that the galaxies with higher $\sigma_\mathrm{gas}$ have flatter metallicity gradients, suggesting a connection between the strength of feedback and metallicity gradient flattening. The driver of higher $\sigma_\mathrm{gas}$ and the flattening of the metallicity gradient may be the same.} 

Most of the observations of high redshift galaxies showed shallow-to-flat metallicity gradient (Figure \ref{fig:gradient_redshift}), which agrees with higher $\sigma_\mathrm{gas}$ and stronger gas transportation and accretion \reviewer{in} higher redshift galaxies \citep{Wisnioski:2015,Ubler:2019,Mai:2024}. However, the flat gradients at high redshift may result from the limit of resolution and the flattening caused by the seeing \citep{Yuan:2013,Belfiore:2017}, as we discuss in Section \ref{sec:seeing effect}.










\section{Conclusions}\label{sec:conclusions-4}

We measured the seeing-deconvolved gas-phase metallicity of 70 star-forming galaxies at $z\sim 0.3$ from the MAGPI survey. We used a Bayesian inference method, \textsc{blobby3d}, to model the flux distribution of different emission lines. This method takes the effect of seeing into account and has high flexibility with regard to the flux distribution. We calculated the metallicity using the N2O2 and N2H$\alpha$ diagnostics \citep{Kewley:2019}. We summarise the main findings below.

(i) The median N2O2 metallicity gradient of our sample is $-0.013^{+0.059}_{-0.033}$ dex/kpc. Among the 70 galaxies, 32.9\% exhibit significant ($\geq 2\sigma$) negative gradients, 10.0\% show significant positive gradients, and 57.1\% are consistent with flat gradients, the latter likely driven by turbulence and feedback-induced mixing.

(ii) The $\nabla \mathrm{[O/H]}$-$M_*$ relation of the MAGPI galaxies is generally consistent with the theoretical prediction of \citet{Sharda:2024}, \reviewer{which suggests that multiple mechanisms, including stellar feedback, gas transportation and gas accretion, contribute to the shaping of the metallicity profile, and the relative importance of these mechanisms varies with the stellar mass of galaxies.} \reviewer{However, the scatter and uncertainty of metallicity gradients of low-mass galaxies are large in our sample. Further study with a large sample is needed to understand the variation of metallicity gradients at $z\sim 0.3$.}


\paperrevise{(iii) The N2O2 metallicity gradients (measured in units of dex/kpc) are positively correlated with $\sigma_\mathrm{gas}$ ($r=0.36^{+0.05}_{-0.06}$) and $\Sigma_\mathrm{SFR}$ ($r=0.31^{+0.04}_{-0.04}$). The metallicity gradients also have a negative correlation with $R_\mathrm{e}$ ($r=-0.37^{+0.08}_{-0.07}$). Partial correlation analysis shows that $\sigma_\mathrm{gas}$ exhibits the strongest correlation with the metallicity gradients, followed by $R_\mathrm{e}$, while SFR shows no significant correlation. Galaxies with higher gas turbulence tend to have flatter or even positive metallicity gradients. These results suggest that mechanisms driving gas turbulence, such as stellar feedback, gas accretion, and transportation, enhance gas mixing, thereby leading to flattened or inverted metallicity gradients. In addition, smaller galaxies may be more susceptible to the accretion of metal-poor gas, \teamreview{or stellar feedback-driven outflows}, and thus exhibit inverted (positive) metallicity gradients.} 

\paperrevise{(iv) The metallicity gradients measured using N2O2 and N2H$\alpha$ diagnostics are broadly consistent ($r=0.53$). Some galaxies have dips in their radial N2H$\alpha$ metallicity profiles or show elevated central N2H$\alpha$ metallicity compared to N2O2 measurements. The difference between them may result from the variation of the ionisation parameter and ISM pressure.}

\teamreview{(v) Metallicity gradients show only a mild evolution with redshift. The significant scatter observed at all epochs may arise from variations in galaxy mass, size, star formation activity, or environment.}

\paperrevise{(vi) We confirm that the seeing-convolved metallicity gradients are flatter than the seeing-deconvolved metallicity gradients. This effect is mitigated when the size of galaxies is large ($>6\times$) compared to the seeing.}

\teamreview{Overall, our results demonstrate that metallicity gradients at $z\sim0.3$ are shaped by a combination of internal and external processes such as star formation, stellar feedback, gas accretion and transportation. The correlation with gas turbulence highlights the importance of gas mixing processes in redistributing metals. While the $\nabla \mathrm{[O/H]}$-$M_*$ relation suggests that the balance of these mechanisms evolves with stellar mass.}

\teamreview{Future work will aim to expand the galaxy sample, especially at the low-mass end, where the large scatter and uncertainties limit robust conclusions. A larger dataset will improve statistical constraints on metallicity gradients and enable a clearer assessment of the relative roles of stellar feedback, gas accretion, internal gas transport, and environmental effects, and how these processes vary with stellar mass. Future IFU surveys with higher spatial resolution and sensitivity, enabled by upcoming AO-assisted instruments, e.g. the Multi-conjugate Adaptive-optics Visible Imager-Spectrograph \citep[MAVIS;][]{McDermid:2020} on the VLT, the Giant Magellan Telescope integral fields spectrograph and imager \citep[GMTIFS;][]{Sharp:2016}, and the High Angular Resolution Monolithic Optical and Near-infrared Integral field spectrograph \citep[HARMONI;][]{Thatte:2021}, will be essential to push metallicity gradient studies to higher redshifts and smaller spatial scales.}

\section*{Acknowledgements}

We wish to thank the ESO staff, and in particular the staff at Paranal Observatory, for carrying out the MAGPI observations. MAGPI targets were selected from GAMA. GAMA is a joint European-Australasian project based around a spectroscopic campaign using the Anglo-Australian Telescope. GAMA was funded by the STFC (UK), the ARC (Australia), the AAO, and the participating institutions. GAMA photometry is based on observations made with ESO Telescopes at the La Silla Paranal Observatory under programme ID 179.A-2004, ID 177.A-3016. The MAGPI team acknowledge support by the Australian Research Council Centre of Excellence for All Sky Astrophysics in 3 Dimensions (ASTRO 3D), through project number CE170100013. YM is supported by an Australian Government Research Training Program (RTP) Scholarship. LMV acknowledges support by the German Academic Scholarship Foundation (Studienstiftung des deutschen Volkes) and the Marianne-Plehn-Program of the Elite Network of Bavaria. I.B. has received funding from the European Union's Horizon 2020 research and innovation programme under the Marie Sklodowska-Curie Grant agreement ID n.º 101059532. This project was extended for 6 months by the Franziska Seidl Funding Program of the University of Vienna. SMS acknowledges funding from the Australian Research Council (DE220100003). Parts of this research were conducted by the Australian Research Council Centre of Excellence for All Sky Astrophysics in 3 Dimensions (ASTRO 3D), through project number CE170100013. KH acknowledges support by the Royal Society through a Dorothy Hodgkin Fellowship to KA Oman (DHF/R1/231105). TG acknowledges support from ARC Discovery Project DP210101945. CF is the recipient of an Australian Research Council Future Fellowship (project number FT210100168) funded by the Australian Government. 
CF is a recipient of ARC Discovery Project DP210101945. 
PS is supported by the Leiden University Oort Fellowship and the International Astronomical Union -- Gruber Foundation (TGF) Fellowship.

\section*{Data Availability}

The MUSE data used in this work are available on the ESO public archive. The reduced MAGPI datacubes and the emission line data products will be public in the MAGPI team data release (Mendel et al. in preparation and Battisti et al. in preparation).


\bibliographystyle{mnras}
\bibliography{mnras_metal_gradient_MAGPI_blobby3d} 




\appendix

\section{\textsc{blobby3d} emission-line flux maps}\label{sec:emission line flux}

\begin{figure*}
    \centering
    \includegraphics [width=0.85\textwidth]{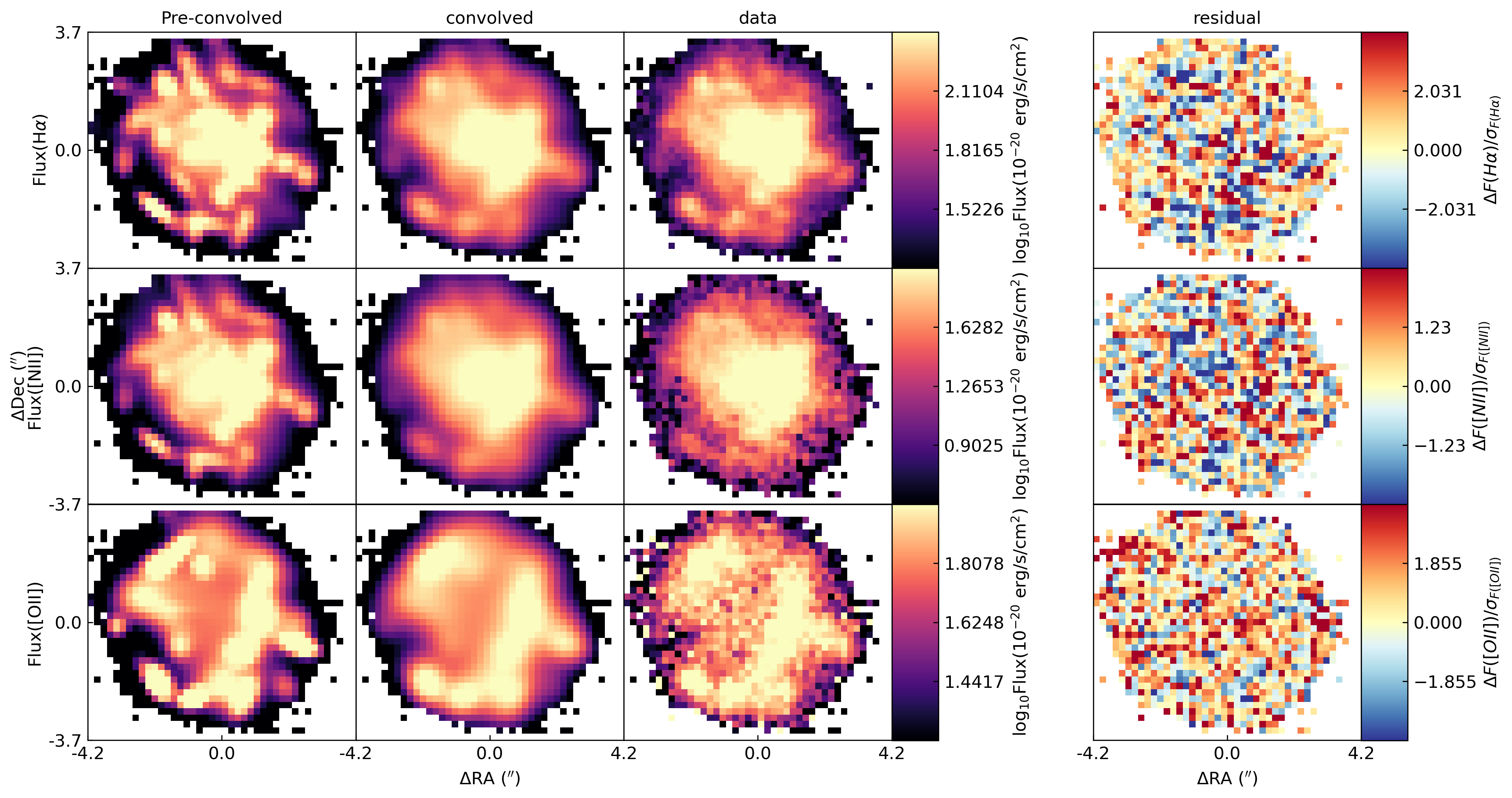}
    \caption{\mnras{The 2D flux maps from the \textsc{blobby3d} model and observational data for MAGPI galaxy 1203195161. The first column shows the seeing-deconvolved flux distribution from the model, the second column shows the corresponding seeing-convolved model flux, the third column shows the flux map obtained from a single-Gaussian fit to the observed data cube, and the fourth column shows the residuals between the observed data and the seeing-convolved model. The first, second and third rows correspond to the H$\alpha$, [N\,\textsc{ii}] and [O\,\textsc{ii}] lines, respectively.}}
    \label{fig:fluxmap_1203}
\end{figure*}

\begin{figure*}
    \centering
    \includegraphics [width=0.85\textwidth]{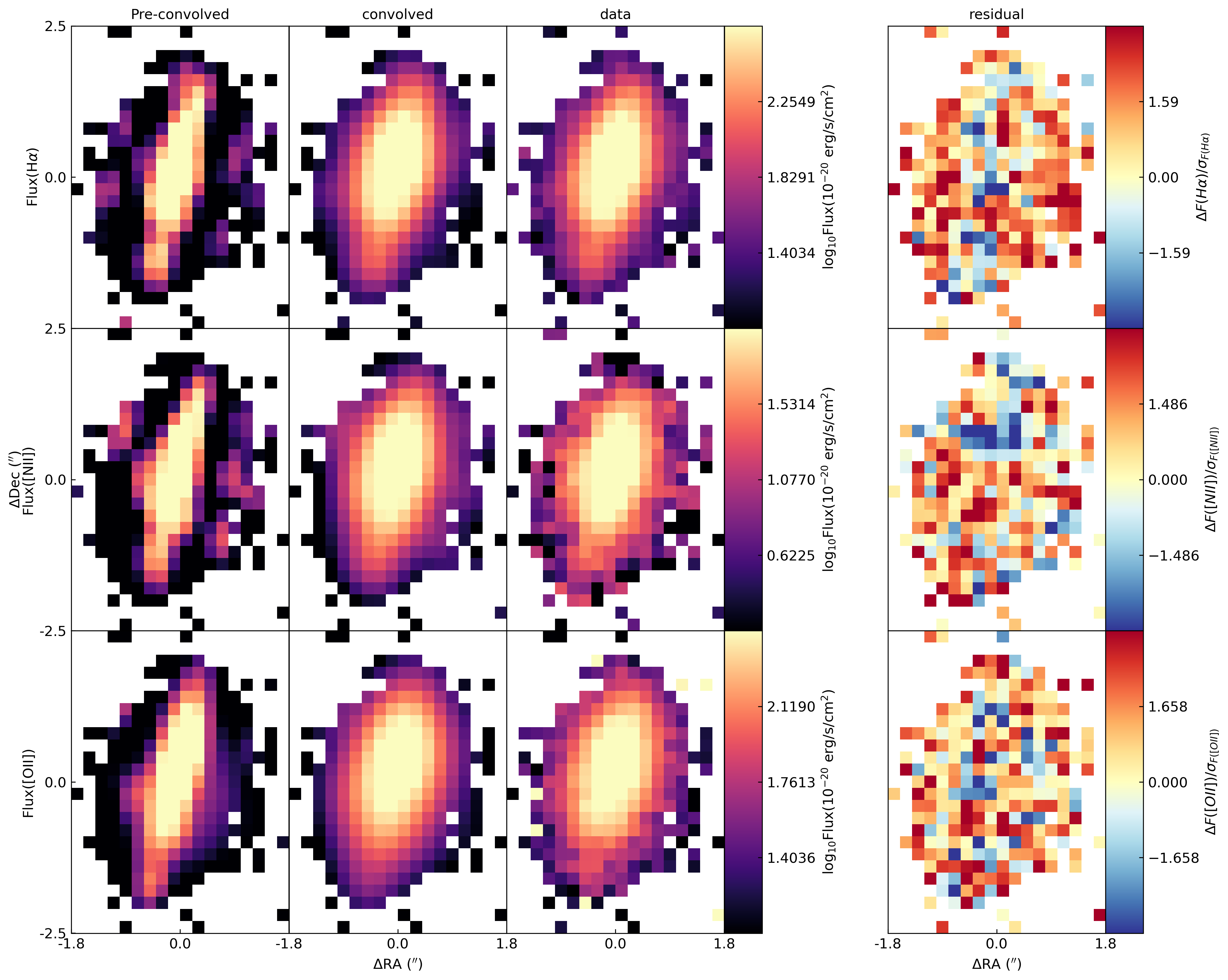}
    \caption{\mnras{Same as Figure \ref{fig:fluxmap_1203}, but for MAGPI galaxy 1508217276.}}
    \label{fig:fluxmap_1508}
\end{figure*}

\mnras{We show the 2D flux maps of the \textsc{blobby3d} results for two MAGPI galaxies in our sample (MAGPIID 1203195161, MAGPIID 1508217276) in Figure \ref{fig:fluxmap_1203} and Figure \ref{fig:fluxmap_1508}. These maps illustrate the emission-line flux distributions derived from the \textsc{blobby3d} forward-modelling process, which simultaneously fits the kinematics and spatial distribution of each line while accounting for the effects of seeing.}

\section{Compiled data}\label{sec:compiled data}

\teamreview{We list the metallicity gradient and central metallicity measured using the N2O2 and N2H$\alpha$ diagnostics, and galaxy properties for 70 galaxies in our sample in Table \ref{tab:metal_table_p1}.}

\begin{center}
\begin{table*}
\centering
\caption{(1) Target name, (2-3) Metallicity gradient and central metallicity measured using the N2O2 diagnostic, (4-5) Metallicity gradient and central metallicity measured using the N2H$\alpha$ diagnostic, (6) Stellar mass, (7) Star formation rate, (8) Intrinsic ionised gas velocity dispersion taken from \citet{Mai:2024}, (9) Effective radius measured from $i$-band image, (10) Redshift.}
\begin{tabular}{crrrrrrrrr}
\hline
MAGPIID & gradient (N2O2) & central (N2O2) & gradient (N2H$\alpha$) & central (N2H$\alpha$) & $\log_{10} M_*$ & SFR & $\sigma_\mathrm{gas}$ & $R_e$ & $z$ \\
 & [dex/kpc] & 12+log$_{10}$(O/H) & [dex/kpc] & 12+log$_{10}$(O/H) & [M$_\odot$] &  [M$_\odot$/yr] & [km/s] & [kpc] &   \\
\hline
1201302222 & $-0.108\pm0.050$ & $8.374\pm0.117$ & $-0.023\pm0.021$ & $8.280\pm0.015$ &$9.23$ &$0.36$ &$16.89$ &$2.85$ &$0.30$ \\
1202077074 & $-0.013\pm0.024$ & $8.645\pm0.049$ & $-0.010\pm0.021$ & $8.461\pm0.033$ &$9.28$ &$0.59$ &$25.94$ &$5.04$ &$0.40$ \\
1203076068 & $-0.009\pm0.008$ & $8.821\pm0.036$ & $-0.021\pm0.004$ & $8.615\pm0.025$ &$10.33$ &$4.79$ &$43.22$ &$4.20$ &$0.31$ \\
1203191233 & $0.011\pm0.011$ & $8.582\pm0.046$ & $-0.022\pm0.001$ & $8.461\pm0.007$ &$9.94$ &$0.33$ &$10.96$ &$6.52$ &$0.35$ \\
1203195161 & $-0.017\pm0.001$ & $8.996\pm0.005$ & $-0.018\pm0.003$ & $8.806\pm0.022$ &$10.72$ &$2.77$ &$22.67$ &$6.60$ &$0.28$ \\
1203235348 & $-0.018\pm0.029$ & $8.798\pm0.063$ & $-0.016\pm0.023$ & $8.496\pm0.035$ &$9.79$ &$0.54$ &$21.13$ &$4.58$ &$0.36$ \\
1203247089 & $-0.004\pm0.042$ & $8.063\pm0.116$ & $0.003\pm0.020$ & $8.218\pm0.003$ &$8.76$ &$0.35$ &$23.37$ &$3.84$ &$0.31$ \\
1205196165 & $0.001\pm0.003$ & $8.919\pm0.012$ & $-0.003\pm0.004$ & $8.691\pm0.016$ &$9.67$ &$0.22$ &$15.73$ &$2.75$ &$0.29$ \\
1206151090 & $-0.021\pm0.007$ & $8.785\pm0.019$ & $-0.020\pm0.003$ & $8.557\pm0.009$ &$9.63$ &$2.13$ &$35.38$ &$3.48$ &$0.32$ \\
1206322202 & $-0.084\pm0.026$ & $8.454\pm0.042$ & $-0.043\pm0.022$ & $8.368\pm0.024$ &$9.88$ &$0.40$ &$23.83$ &$2.54$ &$0.33$ \\
1207181305 & $-0.032\pm0.008$ & $8.751\pm0.023$ & $-0.018\pm0.002$ & $8.609\pm0.006$ &$10.37$ &$0.99$ &$30.66$ &$4.57$ &$0.32$ \\
1208109039 & $-0.015\pm0.003$ & $9.003\pm0.019$ & $-0.007\pm0.002$ & $8.606\pm0.017$ &$10.52$ &$1.96$ &$25.02$ &$6.68$ &$0.37$ \\
1501329201 & $0.019\pm0.032$ & $7.989\pm0.078$ & $-0.013\pm0.020$ & $8.222\pm0.013$ &$7.78$ &$0.11$ &$16.92$ &$4.20$ &$0.27$ \\
1502071104 & $-0.002\pm0.027$ & $8.799\pm0.035$ & $-0.006\pm0.020$ & $8.599\pm0.005$ &$10.03$ &$1.29$ &$58.57$ &$4.00$ &$0.37$ \\
1502079084 & $0.056\pm0.036$ & $8.461\pm0.067$ & $-0.010\pm0.022$ & $8.405\pm0.017$ &$9.52$ &$0.61$ &$17.53$ &$2.04$ &$0.34$ \\
1502275236 & $0.185\pm0.043$ & $7.541\pm0.090$ & $-0.000\pm0.020$ & $8.030\pm0.000$ &$8.47$ &$0.57$ &$33.12$ &$0.92$ &$0.32$ \\
1503316290 & $0.017\pm0.037$ & $8.668\pm0.105$ & $-0.006\pm0.024$ & $8.508\pm0.044$ &$9.93$ &$0.33$ &$17.92$ &$3.35$ &$0.38$ \\
1504220116 & $-0.011\pm0.006$ & $8.888\pm0.023$ & $-0.019\pm0.006$ & $8.637\pm0.027$ &$9.77$ &$0.47$ &$23.15$ &$3.94$ &$0.31$ \\
1504328073 & $0.024\pm0.025$ & $8.324\pm0.048$ & $-0.002\pm0.020$ & $8.368\pm0.004$ &$9.13$ &$0.58$ &$27.71$ &$3.84$ &$0.38$ \\
1506082320 & $0.023\pm0.025$ & $8.272\pm0.041$ & $-0.002\pm0.020$ & $8.302\pm0.003$ &$9.67$ &$1.02$ &$31.21$ &$1.93$ &$0.29$ \\
1506106169 & $-0.030\pm0.002$ & $8.991\pm0.007$ & $-0.022\pm0.002$ & $8.802\pm0.012$ &$10.15$ &$0.73$ &$25.06$ &$3.25$ &$0.29$ \\
1507211057 & $-0.045\pm0.023$ & $8.987\pm0.022$ & $-0.002\pm0.022$ & $8.679\pm0.019$ &$10.10$ &$1.25$ &$17.70$ &$2.03$ &$0.32$ \\
1508217276 & $-0.006\pm0.012$ & $8.599\pm0.035$ & $-0.011\pm0.004$ & $8.496\pm0.013$ &$9.85$ &$1.27$ &$26.90$ &$3.75$ &$0.30$ \\
1508336146 & $0.029\pm0.029$ & $8.491\pm0.046$ & $-0.009\pm0.021$ & $8.408\pm0.017$ &$9.39$ &$0.17$ &$18.34$ &$2.52$ &$0.33$ \\
1509152248 & $-0.062\pm0.027$ & $8.727\pm0.052$ & $-0.035\pm0.021$ & $8.454\pm0.029$ &$8.99$ &$0.16$ &$17.19$ &$4.17$ &$0.39$ \\
1509257282 & $-0.029\pm0.022$ & $8.906\pm0.019$ & $-0.002\pm0.020$ & $8.367\pm0.001$ &$10.07$ &$0.06$ &$31.11$ &$7.16$ &$0.42$ \\
1509291230 & $0.062\pm0.034$ & $8.234\pm0.097$ & $-0.007\pm0.001$ & $8.340\pm0.002$ &$10.29$ &$1.02$ &$33.03$ &$2.89$ &$0.30$ \\
1511160258 & $-0.033\pm0.005$ & $8.943\pm0.021$ & $-0.029\pm0.003$ & $8.723\pm0.015$ &$10.09$ &$0.54$ &$18.55$ &$5.28$ &$0.29$ \\
1512115127 & $-0.069\pm0.009$ & $8.820\pm0.026$ & $-0.048\pm0.003$ & $8.615\pm0.011$ &$9.08$ &$0.26$ &$17.80$ &$3.41$ &$0.32$ \\
1513187164 & $-0.022\pm0.020$ & $8.916\pm0.006$ & $0.015\pm0.023$ & $8.581\pm0.016$ &$9.77$ &$0.69$ &$25.56$ &$3.67$ &$0.32$ \\
1513284105 & $0.005\pm0.050$ & $8.304\pm0.106$ & $-0.001\pm0.020$ & $8.266\pm0.001$ &$9.16$ &$0.50$ &$31.42$ &$1.81$ &$0.32$ \\
1514079248 & $-0.027\pm0.021$ & $8.518\pm0.077$ & $-0.015\pm0.002$ & $8.386\pm0.010$ &$9.38$ &$0.20$ &$19.54$ &$4.00$ &$0.30$ \\
1516091250 & $0.046\pm0.021$ & $8.472\pm0.011$ & $-0.019\pm0.021$ & $8.463\pm0.007$ &$9.31$ &$0.44$ &$40.64$ &$1.65$ &$0.37$ \\
1517143230 & $-0.007\pm0.001$ & $8.977\pm0.007$ & $0.006\pm0.004$ & $8.636\pm0.027$ &$10.68$ &$1.66$ &$22.96$ &$5.48$ &$0.29$ \\
1517159274 & $-0.033\pm0.008$ & $8.940\pm0.033$ & $-0.032\pm0.003$ & $8.719\pm0.016$ &$10.10$ &$2.63$ &$28.53$ &$5.42$ &$0.41$ \\
1517265320 & $-0.079\pm0.006$ & $8.940\pm0.018$ & $-0.035\pm0.003$ & $8.569\pm0.020$ &$9.39$ &$0.59$ &$21.21$ &$5.13$ &$0.30$ \\
1519062299 & $-0.015\pm0.027$ & $8.137\pm0.170$ & $-0.018\pm0.006$ & $8.341\pm0.039$ &$9.71$ &$0.27$ &$25.29$ &$4.15$ &$0.40$ \\
1519164262 & $-0.023\pm0.025$ & $8.582\pm0.059$ & $-0.000\pm0.020$ & $8.384\pm0.013$ &$9.48$ &$0.61$ &$25.68$ &$3.28$ &$0.40$ \\
1523135170 & $0.023\pm0.009$ & $8.551\pm0.034$ & $-0.002\pm0.002$ & $8.456\pm0.008$ &$9.92$ &$1.55$ &$23.22$ &$4.19$ &$0.32$ \\
1524292106 & $-0.145\pm0.032$ & $8.708\pm0.043$ & $-0.035\pm0.021$ & $8.440\pm0.009$ &$8.51$ &$0.22$ &$17.70$ &$2.09$ &$0.33$ \\
1528310241 & $-0.006\pm0.012$ & $8.527\pm0.033$ & $-0.013\pm0.004$ & $8.490\pm0.013$ &$9.72$ &$0.22$ &$22.22$ &$4.39$ &$0.32$ \\
1529336091 & $0.110\pm0.029$ & $8.327\pm0.040$ & $0.000\pm0.020$ & $8.418\pm0.000$ &$9.01$ &$0.32$ &$32.98$ &$1.14$ &$0.29$ \\
1530070238 & $-0.008\pm0.027$ & $8.917\pm0.039$ & $-0.007\pm0.020$ & $8.708\pm0.010$ &$9.83$ &$0.55$ &$26.57$ &$3.15$ &$0.37$ \\
1530093302 & $0.069\pm0.045$ & $7.771\pm0.096$ & $-0.000\pm0.020$ & $8.173\pm0.000$ &$9.32$ &$0.29$ &$25.46$ &$2.54$ &$0.37$ \\
1530272149 & $-0.020\pm0.013$ & $8.814\pm0.044$ & $-0.017\pm0.001$ & $8.636\pm0.006$ &$9.87$ &$0.08$ &$16.68$ &$4.72$ &$0.31$ \\
1530322331 & $-0.004\pm0.003$ & $8.863\pm0.019$ & $-0.002\pm0.004$ & $8.685\pm0.025$ &$11.09$ &$3.51$ &$50.32$ &$5.60$ &$0.37$ \\
1531103073 & $-0.088\pm0.040$ & $8.542\pm0.073$ & $-0.012\pm0.022$ & $8.395\pm0.025$ &$9.63$ &$0.16$ &$14.88$ &$3.49$ &$0.35$ \\
1531178256 & $-0.014\pm0.021$ & $8.958\pm0.021$ & $-0.000\pm0.020$ & $8.582\pm0.005$ &$10.31$ &$0.31$ &$18.82$ &$3.35$ &$0.35$ \\
1532240333 & $0.062\pm0.033$ & $7.877\pm0.103$ & $-0.012\pm0.021$ & $8.237\pm0.017$ &$9.07$ &$0.38$ &$24.64$ &$4.79$ &$0.38$ \\
1532282113 & $0.066\pm0.029$ & $8.566\pm0.055$ & $-0.001\pm0.021$ & $8.417\pm0.014$ &$9.82$ &$1.41$ &$30.18$ &$3.52$ &$0.28$ \\
1533161121 & $-0.032\pm0.002$ & $8.939\pm0.007$ & $-0.024\pm0.004$ & $8.684\pm0.019$ &$9.80$ &$0.82$ &$23.64$ &$4.78$ &$0.41$ \\
1534070145 & $-0.049\pm0.030$ & $8.746\pm0.041$ & $-0.019\pm0.023$ & $8.605\pm0.020$ &$10.16$ &$0.89$ &$21.94$ &$1.95$ &$0.31$ \\
1534194077 & $-0.006\pm0.003$ & $8.952\pm0.012$ & $0.004\pm0.009$ & $8.647\pm0.035$ &$10.49$ &$4.24$ &$25.32$ &$3.62$ &$0.31$ \\
1534261263 & $-0.046\pm0.026$ & $8.318\pm0.053$ & $-0.019\pm0.020$ & $8.306\pm0.015$ &$9.19$ &$0.30$ &$18.19$ &$3.52$ &$0.37$ \\
1534282147 & $0.052\pm0.027$ & $8.010\pm0.064$ & $-0.003\pm0.020$ & $8.302\pm0.005$ &$8.67$ &$0.26$ &$41.00$ &$3.11$ &$0.32$ \\
1535157300 & $-0.023\pm0.022$ & $7.806\pm0.032$ & $-0.013\pm0.020$ & $8.162\pm0.014$ &$9.35$ &$0.58$ &$25.91$ &$3.90$ &$0.41$ \\
2301064121 & $0.002\pm0.020$ & $8.916\pm0.013$ & $0.032\pm0.022$ & $8.520\pm0.028$ &$10.73$ &$2.73$ &$32.66$ &$4.13$ &$0.29$ \\
2301109255 & $-0.005\pm0.002$ & $8.931\pm0.009$ & $-0.001\pm0.007$ & $8.679\pm0.036$ &$10.04$ &$2.51$ &$23.82$ &$4.74$ &$0.29$ \\
2302280234 & $0.125\pm0.044$ & $8.161\pm0.078$ & $0.000\pm0.020$ & $8.290\pm0.000$ &$7.85$ &$0.32$ &$40.67$ &$0.92$ &$0.29$ \\
2304104201 & $-0.026\pm0.003$ & $8.987\pm0.017$ & $-0.016\pm0.003$ & $8.698\pm0.034$ &$10.58$ &$5.86$ &$25.45$ &$5.59$ &$0.29$ \\
\hline
\end{tabular}
\label{tab:metal_table_p1}
\end{table*}
\end{center}

\begin{center}
\begin{table*}
\centering
\contcaption{}
\begin{tabular}{crrrrrrrrr}
\hline
MAGPIID & gradient (N2O2) & central (N2O2) & gradient (N2H$\alpha$) & central (N2H$\alpha$) & $\log_{10} M_*$ & SFR & $\sigma_\mathrm{gas}$ & $R_e$ & $z$ \\
 & [dex/kpc] & 12+log$_{10}$(O/H) & [dex/kpc] & 12+log$_{10}$(O/H) & [M$_\odot$] &  [M$_\odot$/yr] & [km/s] & [kpc] &   \\
\hline
2304202299 & $0.156\pm0.050$ & $8.009\pm0.065$ & $0.000\pm0.021$ & $8.258\pm0.000$ &$8.91$ &$0.63$ &$25.57$ &$1.78$ &$0.34$ \\
2305332151 & $-0.025\pm0.024$ & $8.335\pm0.114$ & $-0.003\pm0.021$ & $8.313\pm0.001$ &$9.61$ &$0.09$ &$18.36$ &$3.84$ &$0.31$ \\
2305342325 & $-0.068\pm0.008$ & $8.959\pm0.036$ & $-0.023\pm0.004$ & $8.550\pm0.037$ &$9.60$ &$0.36$ &$21.19$ &$5.70$ &$0.37$ \\
2308076271 & $-0.042\pm0.011$ & $8.925\pm0.017$ & $-0.032\pm0.001$ & $8.663\pm0.018$ &$9.79$ &$0.65$ &$20.04$ &$5.01$ &$0.39$ \\
2308129329 & $-0.068\pm0.001$ & $8.534\pm0.068$ & $-0.028\pm0.003$ & $8.361\pm0.046$ &$9.41$ &$0.45$ &$30.49$ &$4.22$ &$0.39$ \\
2308186140 & $-0.046\pm0.029$ & $8.316\pm0.064$ & $-0.018\pm0.023$ & $8.306\pm0.010$ &$9.14$ &$0.86$ &$26.53$ &$3.34$ &$0.35$ \\
2310037140 & $0.008\pm0.042$ & $8.864\pm0.004$ & $0.006\pm0.020$ & $8.704\pm0.011$ &$9.72$ &$8.02$ &$43.66$ &$1.41$ &$0.34$ \\
2310137110 & $0.054\pm0.003$ & $8.099\pm0.032$ & $0.002\pm0.004$ & $8.224\pm0.004$ &$8.24$ &$0.12$ &$22.03$ &$1.28$ &$0.28$ \\
2310293123 & $-0.198\pm0.007$ & $8.563\pm0.256$ & $-0.023\pm0.003$ & $8.280\pm0.013$ &$8.27$ &$0.08$ &$14.49$ &$1.97$ &$0.28$ \\
2311102063 & $-0.033\pm0.026$ & $8.918\pm0.012$ & $-0.036\pm0.022$ & $8.733\pm0.036$ &$9.76$ &$1.13$ &$29.10$ &$2.61$ &$0.37$ \\
\hline
\end{tabular}
\end{table*}
\end{center}






\bsp	
\label{lastpage}
\end{document}